\documentclass[lettersize,journal]{IEEEtran}
\usepackage{amsmath,amsfonts,bm,amssymb}
\usepackage{array}
\usepackage{textcomp}
\usepackage{url}
\usepackage{verbatim}
\usepackage{graphicx}
\usepackage{cite}
\usepackage{mathtools,nccmath}
\usepackage[ruled,vlined]{algorithm2e}
\usepackage[normalem]{ulem}
\newcommand{\stkout}[1]{\ifmmode\text{\sout{\ensuremath{#1}}}\else\sout{#1}\fi}
\usepackage{cancel}
\usepackage{subcaption}
\usepackage[T1]{fontenc}
\usepackage{xcolor}

\begin{document}
\title{EMORF/S: EM-Based Outlier-Robust Filtering and Smoothing With Correlated Measurement Noise }
	
\author{Aamir Hussain Chughtai, Muhammad Tahir, \IEEEmembership{Senior Member, IEEE}, and Momin Uppal, \IEEEmembership{Senior Member, IEEE}
	\thanks{The authors are with Department of Electrical Engineering, Lahore University of Management Sciences, DHA Lahore Cantt., 54792, Lahore Pakistan. { (email: chughtaiah@gmail.com; tahir@lums.edu.pk; momin.uppal@lums.edu.pk)}}%
}
	
	
	
	\maketitle
	
	\begin{abstract}
In this article, we consider the problem of outlier-robust state estimation where the measurement noise can be correlated. Outliers in data arise due to many reasons like sensor malfunctioning, environmental behaviors, communication glitches, etc. Moreover, noise correlation emerges in several real-world applications e.g. sensor networks, radar data, GPS-based systems, etc. We consider these effects in system modeling which is subsequently used for inference. We employ the Expectation-Maximization (EM) framework to derive both outlier-resilient filtering and smoothing methods, suitable for online and offline estimation respectively. The standard Gaussian filtering and the Gaussian Rauch–Tung–Striebel (RTS) smoothing results are leveraged to devise the estimators. In addition, Bayesian Cramer-Rao Bounds (BCRBs) for a filter and a smoother which can perfectly detect and reject outliers are presented. These serve as useful theoretical benchmarks to gauge the error performance of different estimators. Lastly, different numerical experiments, for an illustrative target tracking application, are carried out that indicate performance gains compared to similarly engineered state-of-the-art outlier-rejecting state estimators. The advantages are in terms of simpler implementation, enhanced estimation quality, and competitive computational performance.

	\end{abstract}
	
	\begin{IEEEkeywords}
		State-Space Models, Approximate Bayesian Inference, Nonlinear Filtering and Smoothing, Outliers, Kalman Filters, Variational Inference, Expectation-Maximization, Robust Estimation, Statistical Learning, Stochastic Dynamical Systems.  
	\end{IEEEkeywords}

\section{Introduction}
\IEEEPARstart{S}{tate} estimation is a key fundamental task in analyzing different dynamical systems with subsequent decision-making and control actions arising in a variety of fields including cybernetics, robotics, power systems, sensor fusion, positioning, and target tracking \cite{9827982,9257077,9643420,9772715,9454284,8684917,jahja2019kalman,9563227,9239326,9050910,9945859} etc. The states describing the system dynamics can evolve intricately. Moreover, these are not directly observable only manifesting themselves in the form of external measurements. Mathematically, it means that for state estimation in general, the inference is performed considering stochastic nonlinear equations making it a nontrivial task.  

\textit{Filtering} is the common term used for online state estimation where inference is carried out at each arriving sample. Kalman filter with its linear and nonlinear versions \cite{kalman1960new,grewal2014kalman,julier1997new} are considered the primary choices for filtering given their ease of implementation and estimation performance. Other options for nonlinear filtering are also available including methods based on Monte-Carlo (MC) approximations e.g. Particle Filters (PFs) \cite{gordon1993novel}, ensemble Kalman filter (EnKF) \cite{evensen2003ensemble} etc. 

\textit{Smoothing}, on the other hand, refers to offline state estimation where the primary concern is not to work on a per-sample basis. We are rather interested in state inference considering the entire batch of measurements. Different options for smoothing exist including the famous Rauch–Tung–Striebel (RTS), two-filter smoothers \cite{rauch1965maximum,sarkka2023bayesian} etc. 

The standard state estimators are devised with the assumption that the dynamical system under consideration is perfectly modeled. The estimators assume the availability of system and observation mathematical models including the process and measurement noise statistics. However, any modeling mismatch can result in deteriorated performance even possibly crippling the functionality of the regular estimators completely.   

In this work, we are interested in coping with the modeling discrepancy and the associated estimation degradation that results from the occurrence of outliers in the measurements. Data outliers can arise due to several factors including data communication problems, environmental variations, and effects, data preprocessing front-end malfunctioning, inherent sensor defects, and degradation, etc \cite{9239326}. We keep our consideration generic by taking into account the possibility of correlated measurement noise with a fully enumerated nominal noise covariance matrix. This is in contrast to the existing approaches where noise in each data dimension is assumed to be independent targeting a specific class of applications \cite{chughtai2022outlier}. {Though the resulting algorithmic structure is endowed with nice computational properties, several important application scenarios where measurement noise correlation exists are not covered. For example, when double differencing of the measurements is performed to get rid of nuisance parameters in Real Time Kinematic (RTK) systems, noise correlation appears in the resulting observation noise vector \cite{9286419}. Likewise, considering upsweep chirp waveforms, a significant negative correlation exists between the range and range rate measurement noise in radar data \cite{953264}.} Similarly, due to the use of a common reference sensor to extract the time difference of arrival (TDOA), correlated range measurement noise arises \cite{5977569}. Besides, in different sensor networks, correlated observation noise also emerges \cite{827356,1549547,4358484}.  
\begin{figure*}[t!]
	\begin{subfigure}{.5\textwidth}
		\centering
		\includegraphics[width=.85\linewidth]{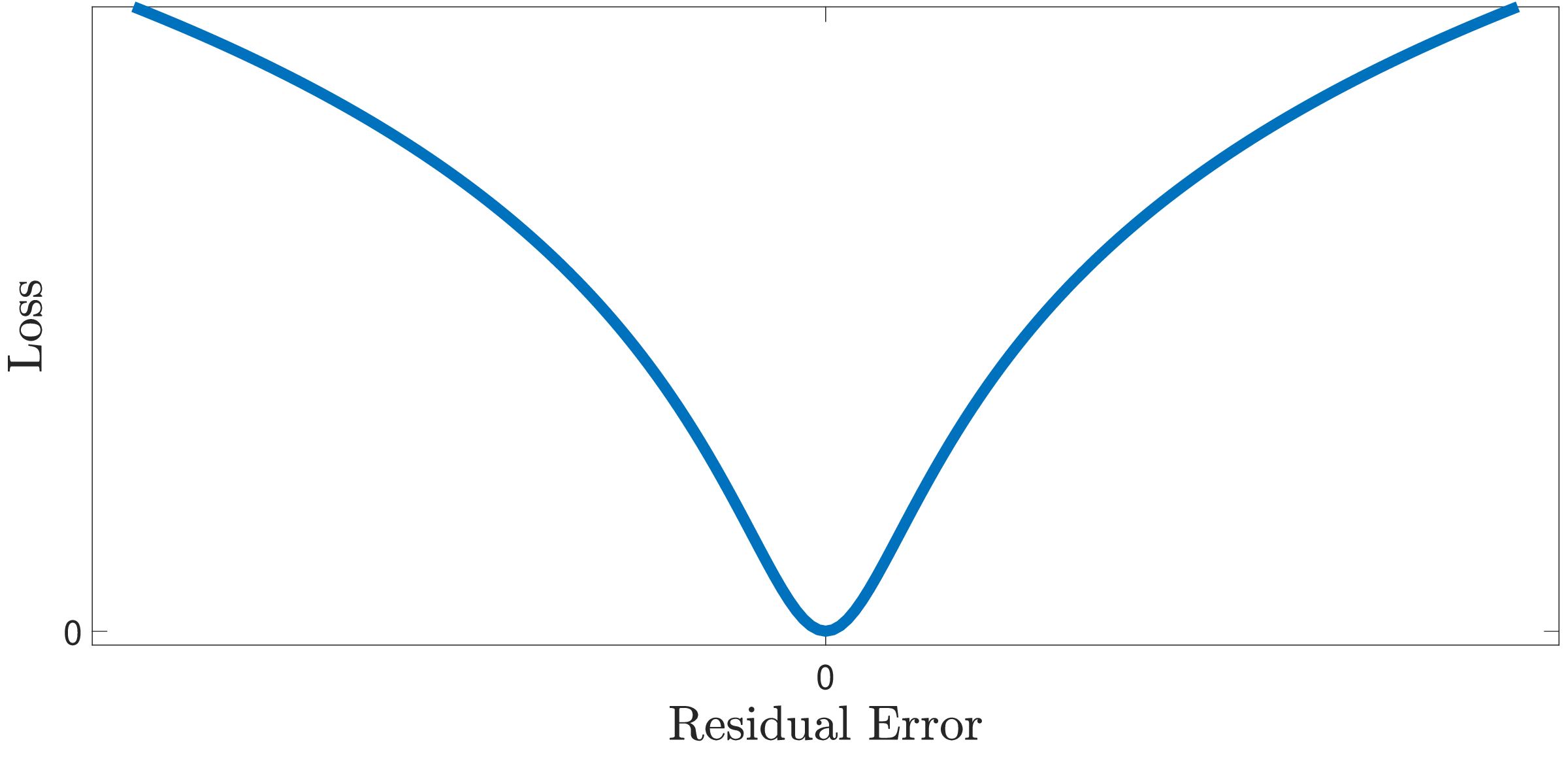}  
		\caption{Static loss functions in traditional methods}
		\label{fig:sub-first}
	\end{subfigure}
	\begin{subfigure}{.5\textwidth}
		\centering
		\includegraphics[width=.85\linewidth]{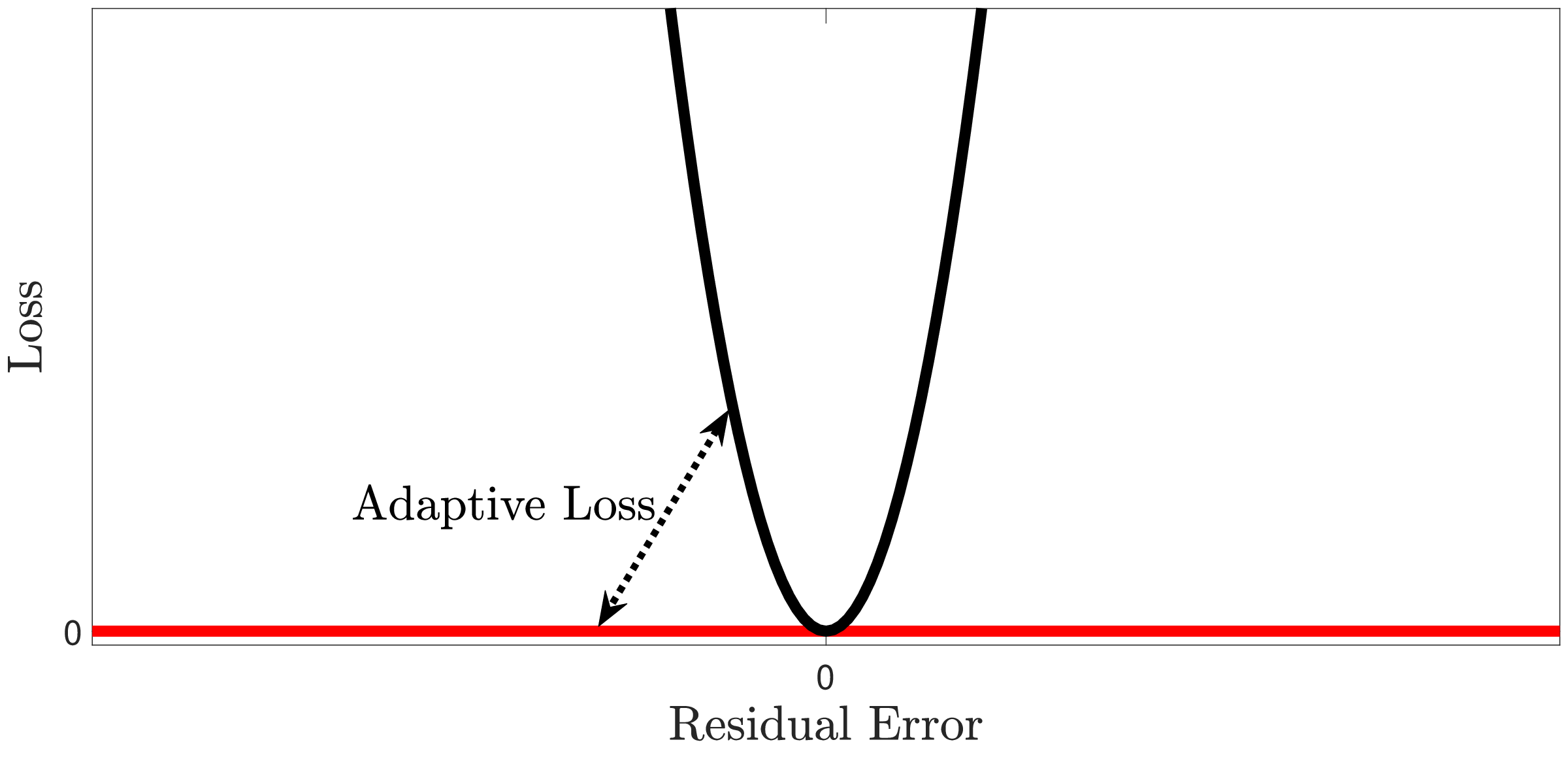}  
		\caption{Adaptive loss functions in learning-based methods}
		\label{fig:sub-second}
	\end{subfigure}
	\caption{Typical loss functions for outlier-robust state estimators. In traditional approaches, the loss function is static. In learning-based methods, the loss function adapts e.g. between a quadratic function and a constant to weight the data during inference.}
	\label{fig:loss}
\end{figure*}

The problem of neutralizing data outliers during state estimation has been approached with various proposals. The traditional way of dealing with outliers is based on assuming fixed statistics for measurement noise or the residuals between predicted and actual measurements. For example, different methods resort to describing observation noise using heavy-tailed distributions like the Student-t and Laplace densities \cite{7812899,8009803}. Similarly, the theory of robust statistics suggests the use of prior models for residuals to downweigh the effect of outliers during inference \cite{karlgaard2015nonlinear,7855662,chang2012multiple,wang2022outlier}. Moreover, some techniques are based on rejecting the data sample by comparing the normalized measurement residuals with some predefined thresholds \cite{6112697,6616007}.

Literature survey indicates that performances of the conventional approaches are sensitive to the tuning of the design parameters which affect the static residual error loss functions during estimation \cite{8869835}. Therefore, tuning-free learning-based techniques have been justified in prior works that make the error loss function adaptive \cite{8869835,8398426,6349794,chughtai2022outlier,9286419,9830121}. These approaches consider appropriate distributions for the measurement noise and subsequently learn the parameters describing the distributions and consequently the loss functions during state estimation. Fig.~\ref{fig:loss} depicts the comparison of typical static loss functions in traditional approaches and dynamic loss functions in learning-based methods considering a uni-dimensional model for visualization (see Section III-D \cite{8869835} for more details). Resultingly, learning-based robust state estimators offer more advantages by reducing user input, being more general, and suiting better for one-shot scenarios. 

Several learning-based methods for robust state estimation have been reported in the relevant literature. As exact inference is not viable for developing these approaches, approximate inference techniques like PFs and variational Bayesian (VB) methods can be used in their design. 
Since PFs can be computationally prohibitive, VB-based techniques are the appealing alternative considering these can leverage the existing standard filtering and smoothing results. Our focus in this work remains on the learning-based outlier mitigation approaches designed using VB.

In previous works, we observe that various outlier-robust state estimators, devised using VB, treat the entire measurement vector collectively during estimation owing to under-parameterized modeling \cite{8869835,8398426,6349794}. Instead of treating each dimension individually, the complete vector is either considered or downweighted by varying the noise covariance matrix by a scalar multiplicative factor. This leaves room for improvement considering useful information is unnecessarily lost during inference. In this regard, we offer a vectorial parameterization to treat each dimension individually in \cite{chughtai2022outlier}. Therein we also suggest a way to make the estimators in \cite{8869835,8398426,6349794} selective. However, these proposals are based on the assumption of independent noise for each measurement dimension. Another learning-based outlier-resilient filter has been presented in \cite{9830121}. But the authors only consider linear systems and test the method with diagonal measurement noise covariance matrices. With the possibility of correlated measurements, the Variational Bayes Kalman Filter (VBKF) has been devised \cite{9286419} by extending the work in \cite{8398426}. However, we observe that VBKF assumes a complex hierarchical model. As a result, along with updating the state densities, it involves updating the nuisance parametric distributions and their hierarchical distributions during the VB updates. This includes evaluation of the digamma function to find the expectation of logarithmic expressions \cite{8398426,9286419}. Therefore, implementing VBKF can get complicated e.g. within an embedded computing device where access to such functions is not inherently available and additional libraries are required. Moreover, extending VBKF to outlier-robust smoothing also gets cumbersome. This calls for simpler state estimation approaches, for systems with correlated noise, that can weather the effect of outliers.  

With this background, considering the possibility of correlated measurements in nonlinear dynamical systems, we make the following contributions in this work. 

\begin{itemize}
	\item Using a suitable model and VB (more specifically Expectation-Maximization (EM)) we devise an outlier-robust filter availing the standard Gaussian filtering results. The results are further utilized in deriving an outlier-robust smoother based on the standard Gaussian RTS smoothing. Since our proposed method is inspired by our prior work which considers independent measurement noise \cite{chughtai2022outlier}, we also present insightful connections.  
	\item We derive Bayesian Cramer-Rao Bounds (BCRBs) for a filter and a smoother which can perfectly detect and reject outliers. This provides a useful benchmark to assess the estimation ability of different outlier-mitigating estimators.
	\item We evaluate the performance of the devised estimators as compared to the other similarly devised outlier-discarding methods. Different scenarios of a relevant TDOA-based target tracking application are considered in numerical experiments indicating the merits of the proposed methods.	  
\end{itemize} 

The rest of the article is organized as follows. Section \ref{Sec_model}
provides the modeling details. In Section \ref{filter}, we present the derivation of the proposed filter. Thereafter, the derivation of the proposed smoother is given in Section \ref{smooth}. In Section \ref{Bounds}, BCRBs for a filter and a smoother with perfect outlier detecting and rejecting capabilities are provided. Subsequently, the performance evaluation results have been discussed in Section \ref{exp}. The paper ends with a conclusive commentary in Section \ref{conc}.

\textit{Notation}: As a general notation in this work, $\mathbf{r}^{\top}$ is the transpose of the vector $\mathbf{r}$, ${{r}^i}$ denotes the $i$th element of a vector $\mathbf{r}$; ${\mathbf{r}^{i-}}$ is the vector $\mathbf{r}$ with its $i$th element removed; the subscript $k$ is used for time index; $\mathbf{r}_{k{}}$ is the vector $\mathbf{r}$ at time instant $k$; $\mathbf{r}_{k{-}}$ is the group of vectors $\mathbf{r}$ considering the entire time horizon except the time instant $k$; ${{R}^{i,j}}$ is the element of the matrix $\mathbf{R}$ present at the $i$th row and $j$th column; $\mathbf{R}^{-1}$ is the inverse of $\mathbf{R}$; $|\mathbf{R}|$ is the determinant of $\mathbf{R}$; $\bm{\mathfrak{R}}$ is the swapped form of $\mathbf{R}$ where the swapping operation is defined in a particular context; $\delta(.)$ represents the delta function; $\langle.\rangle_{q(\bm{\psi}_k)}$ denotes the expectation of the argument with respect to the distribution $q(\bm{\psi}_k)$; $\mathrm{tr}(.)$ is the trace operator; $a\mod b$ denotes the remainder of $a/b$; the superscripts $-$ and $+$ are used for the predicted and updated filtering parameters respectively; the superscript $s$ is used for the parameters of the marginal smoothing densities. Other symbols are defined in their first usage context.

\section{State-space modeling}\label{Sec_model}
\subsection{Standard modeling}
Consider a standard nonlinear discrete-time state-space model (SSM) to represent the dynamics of a physical system given as
\begin{align}
	\mathbf{x}_k&= \mathbf{f}(\mathbf{x}_{k-1})+\mathbf{q}_{k-1}	
	\label{eqn_model_1}\\
	\mathbf{y}_k&= \mathbf{h}(\mathbf{x}_{k})+\mathbf{r}_{k}
	\label{eqn_model_2}
\end{align}
where $\mathbf{x}_k\in \mathbb{R}^n$ and $\textbf{y}_k \in \mathbb{R}^m$ denote the state and measurement vectors respectively; the nonlinear functions  $\textbf{f}(.):\mathbb{R}^n\rightarrow\mathbb{R}^n$ and $\textbf{h}(.):\mathbb{R}^n\rightarrow\mathbb{R}^m$  represent the process dynamics and observation transformations respectively; $\textbf{q}_{k}\in \mathbb{R}^n$ and $\textbf{r}_k\in \mathbb{R}^m$ account for the additive nominal process and measurement noise respectively. $\textbf{q}_{k}$ and $\textbf{r}_k$ are assumed to be statistically independent, White, and normally distributed with zero mean and known covariance matrices $\textbf{Q}_k$ and $\textbf{R}_k$ respectively. We consider that $\textbf{R}_k$ can be a fully enumerated matrix capturing the correlations between the measurement noise entries. 
\subsection{Modeling outliers for inference}
The model in \eqref{eqn_model_1}-\eqref{eqn_model_2} assumes that the measurements are only affected by nominal measurement noise $\textbf{r}_k$. However, the observations in every dimension can be corrupted with outliers leading to the disruption of standard state estimators as the measurement data cannot be described by the regular model. Therefore, data outliers need to be appropriately modeled within the generative SSM with two basic objectives. Firstly, the model should sufficiently capture the effect of outlier contamination in the data. Secondly, the model should remain amenable to inference. 

To model the outliers in SSM, we consider an indicator vector $\bm{\mathcal{I}}_k\in\mathbb{R}^m$ having Bernoulli elements where its $i$th element ${{\mathcal{I}}}^i_k$ can assume two possible values: $\epsilon$ (close to zero) and 1. ${{\mathcal{I}}}^i_k=\epsilon$ denotes the presence of an outlier, whereas ${{\mathcal{I}}}^i_k=1$ indicates no outlier in the $i$th dimension at time $k$. Since outliers can occur independently at any instant, we assume that the elements of $\bm{{\mathcal{I}}}_k$ are statistically independent of their past. Additionally, we assume that the entries of $\bm{{\mathcal{I}}}_k$ to be independent of each other since generally no knowledge of correlations between outliers is available which are not easy to model anyway. Moreover, this choice is motivated by the goal of inferential tractability. We also consider $\bm{\mathcal{I}}_k$ and $\mathbf{x}_k$ to be statistically independent since the outlier occurrence does not depend on the state value. The assumed distribution of $\bm{\mathcal{I}}_k$ is given as
\begin{equation}
	p(\bm{\mathcal{I}}_k)=\prod_{i=1}^{m}p({{\mathcal{I}}}^i_k)=\prod_{i=1}^{m} (1-{\theta^i_k}) \delta({{{\mathcal{I}}}^i_k}-\epsilon)+{\theta^i_k}\delta( {{{\mathcal{I}}}^i_k}-1)
	\label{eqn_model_3}
\end{equation}
where $\theta^i_k$ denotes the prior probability of no outlier in the $i$th observation at time $k$. Further, the conditional measurement likelihood given the current state $\mathbf{x}_k$ and the indicator $\bm{\mathcal{I}}_k$, is proposed to be normally distributed as
\begin{figure}[t!]
	\centering
	\includegraphics[trim={.5cm .5cm .5cm 0.7cm},clip,width=.8\linewidth]{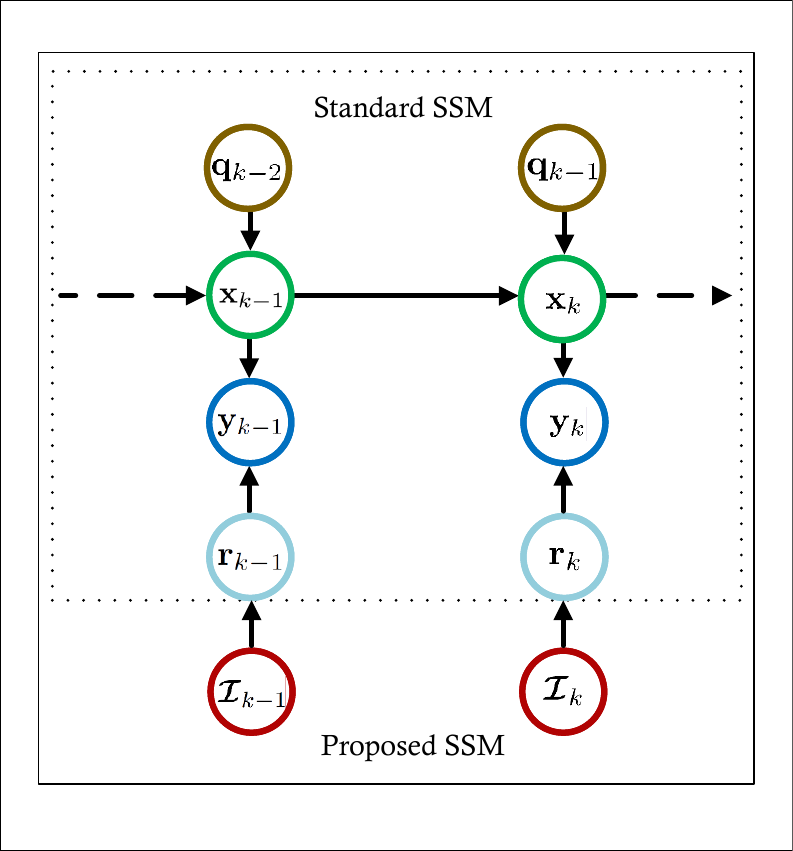}
	\caption{Probabilistic graphical model for the proposed method} 
	\label{fig:PGM}
\end{figure}
\begin{align}
	&p(\mathbf{y}_k|\mathbf{x}_k,\bm{\mathcal{I}}_k)={\mathcal{N}}\big(\mathbf{y}_k|\mathbf{h}(\mathbf{x}_k),\mathbf{R}_k({ \bm{\mathcal{{I}}}_k}) \big)\nonumber\\
	&=\frac{1}{\sqrt{(2 \pi)^{m}|\mathbf{R}_k({ \bm{\mathcal{{I}}}_k})|}}\mathrm{exp} \Big\{ {-}\mfrac{1}{2}{(\mathbf{y}_k-\mathbf{h}(\mathbf{x}_k))}^{\top}\mathbf{R}_k^{-1}(\bm{\mathcal{I}}_k)\nonumber\\
	&\ \ \ \ (\mathbf{y}_k-\mathbf{h}(\mathbf{x}_k)) \Big\}		\label{eqn_model_4}
\end{align}
where $\mathbf{R}_k(\boldsymbol{\mathcal{I}}_{k})$ is defined as
\begin{align}
	 \begin{bmatrix}
	{R}^{1,1}_k / {{{\mathcal{I}}}^1_k}  & \dots &  {R}^{1,m}_k \delta( {{{\mathcal{I}}}^1_k}{-}1) \delta( {{{\mathcal{I}}}^m_k}{-}1) \\
	\vdots &  \ddots& \vdots\\
	{R}^{m,1}_k \delta( {{{\mathcal{I}}}^m_k}{-}1) \delta( {{{\mathcal{I}}}^1_k}{-}1)  & \cdots & {R}^{m,m}_k / {{{\mathcal{I}}}^m_k}
\end{bmatrix} 
\label{eqn_model_5}
\end{align}

$\mathbf{R}_k({ \bm{\mathcal{{I}}}_k})$ is the modified covariance matrix of the measurements considering the effect of outliers. The effect of $\mathbf{R}_k({ \bm{\mathcal{{I}}}_k})$ on the data generation process can be understood by considering the possible values of ${\mathcal{I}}^i_k$. In particular, ${\mathcal{I}}^i_k=\epsilon$ leads to a very large $i$th diagonal entry of $\mathbf{R}_k({ \bm{\mathcal{{I}}}_k})$, while placing zeros at the remaining $i$th row and column of the matrix. Resultingly, when an outlier occurs in the $i$th dimension its effect on state estimation is minimized. Moreover, the $i$th dimension no longer has any correlation with any other entry, ceasing to have any effect on any other dimension during inference. This is in contrast to ${\mathcal{I}}^i_k=1$ which ensures the diagonal element and the off-diagonal correlation entries with other non-affected dimensions are preserved.  Note that the conditional likelihood is independent of the batch of all the historical observations $\mathbf{y}_{1:{k-1}}$. 

{{Considering \eqref{eqn_model_3} and \eqref{eqn_model_4}, we can write the modified measurement model incorporating the effect of outliers as
		\begin{align}
			\mathbf{y}_k&= \mathbf{h}(\mathbf{x}_{k})+\bm{\upsilon}_{k}
			\label{eqn_model_2-m}
		\end{align}
		where the modified measurement noise follows a Gaussian mixture model as $\bm{\upsilon}_{k}\sim \underset{{\bm{\mathcal{I}}_k}}{\sum} \ {\mathcal{N}}(\bm{\upsilon}_k|\mathbf{0},\mathbf{R}_k({ \bm{\mathcal{{I}}}_k}) ) p(\bm{\mathcal{I}}_k)$.  }}
 
Fig.~\ref{fig:PGM} shows how the standard probabilistic graphical model (PGM) is modified into the proposed PGM for devising outlier-robust state estimators. The suggested PGM meets the modeling aims of describing the nominal and corrupted data sufficiently while remaining docile for statistical inference.

\section{Proposed Robust Filtering}\label{filter}
In filtering, we are interested in the posterior distribution of $\mathbf{x}_k$ conditioned on all the observations $\mathbf{y}_{1:{k}}$ that have been observed till time $k$. For this objective, we can employ the Bayes rule recursively. Given the proposed observation model, the analytical expression of the joint posterior distribution of $\mathbf{x}_k$ and $\bm{\mathcal{I}}_k$ conditioned on the set of all the observations $\mathbf{y}_{1:{k}}$ is given as
\begin{equation}
	p(\mathbf{x}_k,\bm{\mathcal{I}}_k|\mathbf{y}_{1:{k}})=\frac{p(\mathbf{y}_k|\bm{\mathcal{I}}_k,\mathbf{x}_{k})	p(\mathbf{x}_k|\mathbf{y}_{1:{k-1}})p(\bm{\mathcal{I}}_k)}{p(\mathbf{y}_k|\mathbf{y}_{1:{k-1}})}
	\label{eqn_fl_1}
\end{equation} 

Theoretically, the joint posterior can further be marginalized to obtain the required posterior distribution $p(\mathbf{x}_k|\mathbf{y}_{1:{k}})$. Assuming $p(\mathbf{x}_k|\mathbf{y}_{1:{k-1}})$ as a  Gaussian distribution, we need to run $2^m$ Kalman filters corresponding to each combination of $\bm{\mathcal{I}}_k$ to obtain the posterior. This results in computational complexity of around $\mathcal{O}(2^{m} m^3)$ where $m^3$ appears due to matrix inversions (ignoring sparsity). Therefore, this approach clearly becomes impractical. 

To get around the problem, we can possibly employ the standard VB method where the product of VB marginals is conveniently used to approximate the joint posterior. We assume the following factorization of the posterior 
\begin{equation}
	p(\mathbf{x}_k,\bm{\mathcal{I}}_k|\mathbf{y}_{1:{k}})\approx q^f(\mathbf{x}_k)\prod_i q^f({\mathcal{I}}^i_k)
	\label{eqn_fl_2}
\end{equation}

The VB approximation aims to minimize the Kullback-Leibler (KL) divergence between the product approximation and the true posterior and leads to the following marginals \cite{murphy2012machine} 
\begin{align}
	q^f(\mathbf{x}_k)&\propto \exp \big( \big\langle\mathrm{ln} ( p(\mathbf{x}_k,\bm{\mathcal{I}}_k|\mathbf{y}_{1:{k}}))\big\rangle_{ q^f({{\bm{\mathcal{I}}}_k})  } \big)\label{eqn_vb_1}\\
	q^f({\mathcal{I}}^i_k)&\propto \exp \big( \big\langle \mathrm{ln} ( p(\mathbf{x}_k,\bm{\mathcal{I}}_k|\mathbf{y}_{1:k}))\big\rangle_{q^f(\mathbf{x}_k)  q^f(\bm{\mathcal{I}}^{i-}_k)  }\ \big) \ \forall\ i \label{eqn_vb_2}
\end{align} 

Using \eqref{eqn_vb_1}-\eqref{eqn_vb_2} alternately the VB marginals can be updated iteratively until convergence. The procedure provides a useful way to approximate the true marginals of the joint posterior by approximating these as $p(\mathbf{x}_k|\mathbf{y}_{1:{k}}) \approx {q^f(\mathbf{x}_k)}$ and $	p(\bm{\mathcal{I}}_k|\mathbf{y}_{1:{k}})\approx \prod_i q^f({\mathcal{I}}^i_k)$.

For our model, \eqref{eqn_vb_1} becomes computationally unfriendly if we use the standard VB approach. In fact, the same complexity order of $\mathcal{O}(m^3 2^m)$ appears as with the basic marginalization approach making this approach intractable too. We elaborate more on it in the upcoming subsection.
\subsection*{Expectation-Maximization as a particular case of variational Bayes}\label{em}
To deal with the complexity issue, instead of considering distributions we can resort to point estimates for ${{\mathcal{I}}}^i_k$. In particular, consider $q^f({\mathcal{I}}^i_k)= \delta({{\mathcal{I}}}^i_k-{\hat{{\mathcal{I}}}}^i_k)$ where ${\hat{{\mathcal{I}}}}^i_k$ denotes the point approximation of ${{{\mathcal{I}}}}^i_k$. Consequently, the variational distributions can be updated in an alternating manner in the Expectation (E) and Maximization (M) steps in the EM algorithm given as \cite{vsmidl2006variational}

\subsubsection*{\textnormal{E-Step}}
\begin{align}
	{q^f(\boldsymbol{\mathbf{x}_k})}&=p(\mathbf{x}_k|\mathbf{y}_{1:{k}},\hat{\bm{\mathcal{I}}}_k)	\propto	p(\mathbf{x}_k,\hat{\bm{\mathcal{I}}}_k|\mathbf{y}_{1:{k}})	\label{eqn_fl_E}
\end{align}
\subsubsection*{\textnormal{M-Step}}
\begin{equation}
	{\hat{\mathcal{I}}}^i_k= \underset{{{\mathcal{I}}}^i_k}{\mathrm{argmax}}\big\langle\mathrm{ln}(p(\mathbf{x}_k,{\mathcal{I}}^i_k,\hat{\bm{\mathcal{I}}}^{i-}_k|\mathbf{y}_{1:{k}})\big\rangle_{q^f(\mathbf{x}_k)} \label{eqn_fl_M}
\end{equation}
where all ${\hat{\mathcal{I}}}^i_k$ in the M-Step are successively updated using the latest estimates. 
\subsection{Prediction}\label{pre_fil}
For filtering, we first obtain the predictive distribution $p(\mathbf{x}_{k}|\mathbf{y}_{1:{k-1}})$ using the posterior distribution at the previous instant $p(\mathbf{x}_{k-1}|\mathbf{y}_{1:{k-1}})$ approximated as Gaussian $q^f(\mathbf{x}_{k-1})\approx {\mathcal{N}}\left(\mathbf{x}_{k-1}|\mathbf{m}^{+}_{k-1},\mathbf{P}^{+}_{k-1}\right)$. Using Gaussian (Kalman) filtering results we make the following approximation \cite{sarkka2023bayesian}
\begin{equation}
	p(\mathbf{x}_k|\mathbf{y}_{1:{k-1}})\approx {\mathcal{N}}\left(\mathbf{x}_k|\mathbf{m}^{-}_k,\mathbf{P}^{-}_{k}\right)\label{eqn_fl_pr0}
\end{equation}
where
\begin{flalign}
	\mathbf{m}^{-}_{k}=&\int \mathbf{f}(\mathbf{x}_{k-1})\  {\mathcal{N}}\left(\mathbf{x}_{k-1}|\mathbf{m}^{+}_{k-1},\mathbf{P}^{+}_{k-1}\right)	d\mathbf{x}_{k-1}&\label{eqn_fl_pr1}\\
	\mathbf{P}^{-}_{k}=&\int\Big\{( \mathbf{f}(\mathbf{x}_{k-1})-\mathbf{m}^{-}_{k})(\mathbf{f}(\mathbf{x}_{k-1})-\mathbf{m}^{-}_{k})^{\top}&\nonumber\\
	&{\mathcal{N}}({\mathbf{x}_{k-1}}|\mathbf{m}^{+}_{k-1},\mathbf{P}^{+}_{k-1})\Big\} d\mathbf{x}_{k-1}+\mathbf{Q}_{k-1}&\label{eqn_fl_pr2}
\end{flalign}
\subsection{Update}\label{upd_fil} 
{Considering the expression of the joint posterior distribution in \eqref{eqn_fl_1}, we can write}
\begin{align}
	\ p\left(\mathbf{x}_{k}, \boldsymbol{\mathcal{I}}_{k}|\mathbf{y}_{1:k}\right) \propto& \mfrac{\mathcal{N}(\mathbf{x}_{k} | \mathbf{m}_{k}^{-}, \mathbf{P}_{k}^{-})}{\sqrt{(2 \pi)^{m}\left|{\mathbf{R}}_{k}(\boldsymbol{\mathcal{I}}_{k})\right|}} \exp \Big\{ {-}\mfrac{1}{2}\left(\mathbf{y}_{k}-\mathbf{h}\left(\mathbf{x}_{k}\right)\right)^{\top}\nonumber\\
	&{\mathbf{R}}_{k}^{-1}(\boldsymbol{\mathcal{I}}_{k})\left(\mathbf{y}_{k}-\mathbf{h}\left(\mathbf{x}_{k}\right)\right)\Big\} \Big\{ \prod_{i} \left(1-\theta_{k}^{i}\right)\nonumber\\
	&\delta\left(\mathcal{I}_{k}^{i}-\epsilon\right)+
	\theta_{k}^{i} \delta\left({\mathcal{I}}_{k}^{i}-1\right) \Big\} \label{eqn_fl_up1_x}
\end{align}
{where we use the expressions of the prior distributions from \eqref{eqn_model_3} and \eqref{eqn_fl_pr0} along with the conditional measurement likelihood in \eqref{eqn_model_4}.}
\subsubsection{Derivation of $q^f(\mathbf{x}_{k})$}
With the E-Step in \eqref{eqn_fl_E} we can write 
\begin{align}
	\ q^f\left(\mathbf{x}_{k}\right) \propto& \exp \Big\{{-}\mfrac{1}{2}\left(\mathbf{y}_{k}-\mathbf{h}\left(\mathbf{x}_{k}\right)\right)^{\top} {\mathbf{R}}_{k}^{-1} (\boldsymbol{\hat{\mathcal{I}}}_{k}) \left(\mathbf{y}_{k}-\mathbf{h}\left(\mathbf{x}_{k}\right)\right) \nonumber\\
	&{-}\mfrac{1}{2}\left(\mathbf{x}_{k}-\mathbf{m}_{k}^{-}\right)^{\top} ({\mathbf{P}_{k}^{-}})^{-1}\left(\mathbf{x}_{k}-\mathbf{m}_{k}^{-}\right)\Big\}\label{eqn_fl_up2_x}
\end{align}
where ${\mathbf{R}}_{k}^{-1}({\boldsymbol{ \hat{\mathcal{I}} }_{k}})$ assumes a particular form resulting from the inversion of ${\mathbf{R}}_{k}({\boldsymbol{ \hat{\mathcal{I}} }_{k}})$ as described in Appendix \ref{FirstAppendix}. 

Note that we avoid evaluating {$\big\langle{ \mathbf{R}}_{k}^{-1} (\boldsymbol{{\mathcal{I}}}_{k}) \big\rangle_{ \Pi_i q^f({\mathcal{I}}^i_k) } $}
that would be required in the standard VB approach. This means that we are able to evade the complexity level of around $\mathcal{O}(m^3 2^m)$ since matrix inversion for each of the $2^m$ combinations are required to evaluate the expectation. However, thanks to EM, we are now working with  ${\mathbf{R}}_{k}^{-1}({\boldsymbol{ \hat{\mathcal{I}} }_{k}})$ which can be evaluated with the maximum complexity of $\mathcal{O}(m^3)$ (considering a fully populated matrix). {Though we lose the preferable uncertainty estimates when resorting to point estimate of $\boldsymbol{{\mathcal{I}}}_{k}$, we pay this cost to devise a practically useful method that is scalable with increase in the measurement dimensionality.} 


To proceed further, we use the results of the general Gaussian filter, to approximate $q^f(\mathbf{x}_k)$ with a Gaussian distribution, $ {\mathcal{N}}\left(\mathbf{x}_k|\mathbf{m}^{+}_k,\mathbf{P}^{+}_{k}\right)$, with parameters updated as
\begin{align}
	\mathbf{m}^{+}_k&=\mathbf{m}^{-}_k+\mathbf{K}_k
	(\mathbf{y}_k-\bm{\mu}_k)	\label{eqn_fl_up4b_x}\\
	\mathbf{P}^{+}_{k}&=\mathbf{P}^{-}_{k}-\mathbf{C}_k\mathbf{K}^\top_k \label{eqn_fl_up5_x}
\end{align}
where
\begin{align}
	\mathbf{K}_k=&\ \mathbf{C}_k (\mathbf{U}_k+{\mathbf{R}}_{k}({\boldsymbol{ \hat{\mathcal{I}} }_{k}}) )^{-1}=\{\mathbf{C}_k ( {\mathbf{R}}_{k}^{-1}({\boldsymbol{ \hat{\mathcal{I}} }_{k}})\nonumber\\ &{-} {\mathbf{R}}_{k}^{-1}({\boldsymbol{ \hat{\mathcal{I}} }_{k}})(\mathbf{I}+\mathbf{U}_k {\mathbf{R}}_{k}^{-1}({\boldsymbol{ \hat{\mathcal{I}} }_{k}}) )^{-1}\mathbf{U}_k {\mathbf{R}}_{k}^{-1}({\boldsymbol{ \hat{\mathcal{I}} }_{k}}) )\}\nonumber\\
	\bm{\mu}_k=&\int \mathbf{h}(\mathbf{x}_k)\  {\mathcal{N}}\left(\mathbf{x}_k|\mathbf{m}^{-}_{k},\mathbf{P}^{-}_{k}\right)	d\mathbf{x}_k\nonumber\\
	\mathbf{U}_k=&\int(\mathbf{h}(\mathbf{x}_k)-\bm{\mu}_k)(\mathbf{h}(\mathbf{x}_k)-\bm{\mu}_k)^{\top}{\mathcal{N}}({\mathbf{x}_k}|\mathbf{m}^{-}_k,\mathbf{P}^{-}_{k})d\mathbf{x}_k\nonumber\\
	\mathbf{C}_k=&\int(\mathbf{x}_k-\mathbf{m}^{-}_{k})(\mathbf{h}(\mathbf{x}_k)-\bm{\mu}_k)^{\top}{\mathcal{N}}({\mathbf{x}_k}|\mathbf{m}^{-}_{k},\mathbf{P}^{-}_{k})d\mathbf{x}_k \nonumber
\end{align}
\subsubsection{Derivation of $\hat{\mathcal{I}}^i_k$}
With the M-Step in \eqref{eqn_fl_M} we can write 
\begin{align}
	{\hat{\mathcal{I}}}^i_k=&\ \underset{{{\mathcal{I}}}^i_k}{\mathrm{argmax}}\label{eqn_fl_up1_i} \big\langle\mathrm{ln}(p(\mathbf{x}_k,{\mathcal{I}}^i_k,\hat{\bm{\mathcal{I}}}^{i-}_k|\mathbf{y}_{1:{k}} )\big\rangle_{q^f(\mathbf{x}_k)}&
\end{align}

Using the Bayes rule we can proceed as
\begin{flalign}
	{\hat{\mathcal{I}}}^i_k=&\  \underset{{{\mathcal{I}}}^i_k}{\mathrm{argmax}}\Big\{ \big\langle\mathrm{ln}(p(\mathbf{y}_{{k}}|\mathbf{x}_{k},{\mathcal{I}}^i_k,\hat{\bm{\mathcal{I}}}^{i-}_k, \stkout{\mathbf{y}_{1:{k-1}}}))\big\rangle_{q^f(\mathbf{x}_{k})}&\nonumber \\
	&+\mathrm{ln} (p({\mathcal{I}}^i_k|\stkout{ {\mathbf{x}_{k}},\hat{\bm{\mathcal{I}}}^{i-}_k,\mathbf{y}_{1:{k-1}} } ))+const. \Big\}& \label{eqn_fl_up2_i}
\end{flalign}
where $const.$ is some constant and $p(\mathbf {x}|\mathbf {y},\stkout {\mathbf {z}})$ denotes the conditional independence of $\mathbf{x}$ and $\mathbf{z}$ given $\mathbf{y}$. {Using the prior expression of $\bm{\mathcal{I}}_k$ in \eqref{eqn_model_3}  and the conditional measurement likelihood in \eqref{eqn_model_4} we can write}
\begin{align}
	{\hat{\mathcal{I}}}^i_k=&\  \underset{{{\mathcal{I}}}^i_k}{\mathrm{argmax}} \Big\{ {-}\mfrac{1}{2} \mathrm{tr} \big( \mathbf{W}_k {\mathbf{R}}_{k}^{-1}( {\mathcal{I}^i_k} , \hat{\bm{\mathcal{I}}}^{i-}_k )     \big) {-}\mfrac{1}{2}	\ln|{\mathbf{R}}_{k}( {\mathcal{I}^i_k} , \hat{\bm{\mathcal{I}}}^{i-}_k ) |\nonumber \\
	&+\ln\big( (1-\theta_{k}^{i})  \delta(\mathcal{I}_{k}^{i}-\epsilon)+\theta_{k}^{i} \delta ({\mathcal{I}}_{k}^{i}-1) \big) \Big\}\label{eqn_fl_up3_i}
\end{align}
where {we use the property of the trace operator applied on the product of matrices given as $\mathrm{tr}(\mathbf{A}\mathbf{B}\mathbf{C})=\mathrm{tr}(\mathbf{C}\mathbf{A}\mathbf{B})$} \cite{petersen2008matrix}. ${\mathbf{R}}_{k}( {\mathcal{I}^i_k} , \hat{\bm{\mathcal{I}}}^{i-}_k )$ denotes ${\mathbf{R}}_{k}({\bm{\mathcal{I}}}_k)$ evaluated at ${\bm{\mathcal{I}}}_k$ with it $i$th element as ${\mathcal{I}^i_k}$ and remaining entries $\hat{\bm{\mathcal{I}}}^{i-}_k $ and 
\begin{align}
	\mathbf{W}_k=\int \left(\mathbf{y}_{k}-\mathbf{h}\left(\mathbf{x}_{k}\right)\right) \left(\mathbf{y}_{k}-\mathbf{h}\left(\mathbf{x}_{k}\right)\right)^{\top} {\mathcal{N}}({\mathbf{x}_k}|\mathbf{m}^{+}_{k},\mathbf{P}^{+}_{k})d\mathbf{x}_k \nonumber
\end{align}

{$\hat{\mathcal{I}}^i_k$ assumes the value $1$ or $\epsilon$ depending on which value maximizes the objective in \eqref{eqn_fl_up3_i} }. Resultingly, $\hat{\mathcal{I}}^i_k$ can be determined as
\begin{align}
	{\hat{\mathcal{I}}}^i_k &= 
	\begin{cases}
		1 & \text{if } \hat{\tau}^i_k \leq 0,\\
		{\epsilon} & \text{if } \hat{\tau}^i_k >0
	\end{cases}\label{If}
\end{align}
with
\begin{flalign}
	\hat{\tau}^i_k =&\ \Big\{ \mathrm{tr} \big( \mathbf{W}_k \triangle{\hat{\mathbf{R}}}_{k}^{-1}     \big) +\ln\Big(\frac{|{\mathbf{R}}_{k}( {\mathcal{I}^i_k=1} , \hat{\bm{\mathcal{I}}}^{i-}_k ) |}{|{\mathbf{R}}_{k}( {\mathcal{I}^i_k=\epsilon} , \hat{\bm{\mathcal{I}}}^{i-}_k )|}\Big)&\nonumber\\
	&+2\ln\Big(\frac{1}{\theta^i_k}-1  \Big) \Big\} & \label{eqn_fl_up5_i}
\end{flalign}
where
\begin{align}
	\triangle{\hat{\mathbf{R}}}_{k}^{-1}=  ({\mathbf{R}}_{k}^{-1}( {\mathcal{I}^i_k=1} , \hat{\bm{\mathcal{I}}}^{i-}_k )-{\mathbf{R}}_{k}^{-1}( {\mathcal{I}^i_k=\epsilon} , \hat{\bm{\mathcal{I}}}^{i-}_k ))\label{eqdelR}
\end{align}

{\eqref{If}-\eqref{eqdelR} allow us to update the estimate of ${\hat{\mathcal{I}}}^i_k$, using the latest estimates of the state and  other dimensions of the indicator vector considering the effect of correlations that can exist in different measurement dimensions.} Using the steps outlined in Appendix \ref{SecondAppendix}, we can further simplify $\hat{\tau}^i_k$ as
\begin{flalign}
	\hat{\tau}^i_k= &\ \Big\{ \mathrm{tr} \big( \mathbf{W}_k \triangle{\hat{\mathbf{R}}}_{k}^{-1}     \big) 
	+\ln \Bigg|\textbf{I}-\frac{\mathbf{R}^{-i,i}_k\mathbf{R}^{i,-i}_k (\hat{\mathbf{R}}^{-i,-i}_{k})^{-1}}{{R}^{i,i}_k}\Bigg| & \nonumber\\
	&+ \ln(\epsilon)+2\ln\Big(\frac{1}{\theta^i_k}-1  \Big) \Big\} &\label{eqn_fl_up10_i}
\end{flalign}
where $\hat{\mathbf{R}}^{-i,-i}_{k}$ is the submatrix of ${\mathbf{R}}_{k}({\boldsymbol{ \hat{\mathcal{I}} }_{k}})$ corresponding to entries of $\hat{\boldsymbol{\mathcal{I}}}^{i-}_k$. $\mathbf{R}^{i,-i}_{k}$ and $\mathbf{R}^{-i,i}_{k}$ contain the measurement covariances between \textit{i}{th} and rest of the dimensions.

\begin{algorithm}[t!]
	\SetAlgoLined
	Initialize\ $\mathbf{m}^{+}_0,\mathbf{P}^{+}_0$\;
	\For{$k=1,2...K$}{
		Initialize $\theta^i_k,{\boldsymbol{ \hat{\mathcal{I}} }_{k}},\mathbf{Q}_k,\mathbf{R}_k$\;
		\textbf{Prediction:} \\
		Evaluate $\mathbf{m}^{-}_k,\mathbf{P}^{-}_k$ with \eqref{eqn_fl_pr1} and \eqref{eqn_fl_pr2}\;
		\textbf{Update:} \\
		\While{not converged}{
			Update ${\mathbf{m}^{+}_k}$ and  ${\mathbf{P}^{+}_k}$ with \eqref{eqn_fl_up4b_x}-\eqref{eqn_fl_up5_x}\;
			Update ${\hat{\mathcal{I}}^i_k}$ $\forall\ i$ with \eqref{If}\;
		}
	}
	\caption{The proposed filter: EMORF}
	\label{Algo1}
\end{algorithm} 

Though we can directly evaluate $\triangle{\hat{\bm{\mathbf{R}}}}_{k}^{-1}$ in \eqref{eqdelR}, we can save computations by avoiding repetitive calculations. To this end, we first need to compute 
\begin{align}
	&\triangle{\hat{\bm{\mathfrak{R}}}}_{k}^{-1}=\begin{bmatrix}
		\bm{\Xi}^{i,i} & \bm{\Xi}^{i,-i} \\
		\bm{\Xi}^{-i,i} & \bm{\Xi}^{-i,-i}  \label{del_t}
	\end{bmatrix}
\end{align}
with
\begin{align}
	\bm{\Xi}^{i,i}&=\frac{1}{R^{i,i}_k-\mathbf{R}^{i,-i}_k(\hat{\mathbf{R}}^{-i,-i}_{k})^{-1} \mathbf{R}^{-i,i}_k}-\frac{\epsilon}{R^{i,i}_k}\label{xi1}\\
	\bm{\Xi}^{i,-i}&=-\frac{\mathbf{R}^{i,-i}_k (\hat{\mathbf{R}}^{-i,-i}_{k})^{-1}}{R^{i,i}_k-\mathbf{R}^{i,-i}_k(\hat{\mathbf{R}}^{-i,-i}_{k})^{-1} \mathbf{R}^{-i,i}_k}\label{xi2}\\
	\bm{\Xi}^{-i,i}&=-\frac{ (\hat{\mathbf{R}}^{-i,-i}_{k})^{-1}\mathbf{R}^{-i,i}_k}{R^{i,i}_k-\mathbf{R}^{i,-i}_k(\hat{\mathbf{R}}^{-i,-i}_{k})^{-1} \mathbf{R}^{-i,i}_k}\label{xi3}\\
	\bm{\Xi}^{-i,-i}&=\frac{ (\hat{\mathbf{R}}^{-i,-i}_{k})^{-1}\mathbf{R}^{-i,i}_k \mathbf{R}^{i,-i}_k (\hat{\mathbf{R}}^{-i,-i}_{k})^{-1}}{R^{i,i}_k-\mathbf{R}^{i,-i}_k(\hat{\mathbf{R}}^{-i,-i}_{k})^{-1} \mathbf{R}^{-i,i}_k}\label{xi4}
\end{align}
where the $i$th row/column entries in $\triangle{\hat{\bm{\mathbf{R}}}}_{k}^{-1}$ have been conveniently swapped with the first row/column elements to obtain $\triangle{\hat{\bm{\mathfrak{R}}}}_{k}^{-1}$. By swapping the first row/column entries of $\triangle{\hat{\bm{\mathfrak{R}}}}_{k}^{-1}$ to the $i$th row/column positions we can reclaim $\triangle{\hat{\bm{\mathbf{R}}}}_{k}^{-1}$. Appendix \ref{ThirdAppendix} provides further details in this regard.

The resulting EM-based outlier-robust filter (EMORF) is outlined as Algorithm \ref{Algo1}. {The Gaussian integrals in EMORF can be approximated using standard numerical integration methods including Gauss-Hermite quadratures, spherical cubature rule, Unscented transformation (UT) and MC techniques \cite{sarkka2023bayesian}.}	

{Note that with the proposed factorization in \eqref{eqn_fl_2}, we obtain a practical and scalable algorithm. The resulting complexity is $\mathcal{O}(m^4)$ following from the evaluation of matrix inverses and determinants for calculating each of the $\hat{\mathcal{I}}^i_k$ $\forall\ i=1\cdots m$ in \eqref{If}.} {Please refer to {Appendix \ref{ZerothAppendix}} for discussion on the issue of complexity in the EM method if we do not force independence between the posterior marginals of ${\mathcal{I}}^i_k$ for inference.} 


{Since the construction of EMORF is inspired from selective observations rejecting filter (SORF) \cite{chughtai2022outlier}, we provide details regarding the connection between these filters in Appendix \ref{ForthAppendix}.}  {As for SORF}, we suggest using the ratio of the L2 norm of the difference of the state estimates from the current and previous VB iterations and the L2 norm of the estimate from the previous iteration {for the convergence criterion in EMORF}. This criterion has been commonly chosen in {other robust filters as well} \cite{9931968,8398426}. {For the prior parameter $\theta^i_{k}$ we suggest choosing a neutral value of 0.5 unless some specific statistical information regarding the rate of outliers is available apriori. For $\epsilon$ we recommend setting its value to be close to zero. For more discussion on the choice of the parameters $\theta^i_{k}$ and $\epsilon$ please refer to Appendix \ref{FifthAppendix}. }

\section{Proposed Robust Smoothing}\label{smooth}
In smoothing, our interest lies in determining the posterior distribution of all the states $\mathbf{x}_{1:K}$ conditioned on the batch of all the observations $\mathbf{y}_{1:{K}}$. With that goal, we take a similar approach to filtering and approximate the joint posterior distribution as a product of marginals
\begin{equation}
	p(\mathbf{x}_{1:K},\bm{\mathcal{I}}_{1:K}|\mathbf{y}_{1:{K}})\approx q^s(\mathbf{x}_{1:K})\prod_{k}\prod_{i} q^s({\mathcal{I}}^i_k)
	\label{eqn_sm_1}
\end{equation}
where the true marginals are approximated as $p(\mathbf{x}_{1:K}|\mathbf{y}_{1:{K}})\approx{q^s(\mathbf{x}_{1:K})}$ and $	p(\bm{\mathcal{I}}_{1:K}|\mathbf{y}_{1:{K}})\approx{\prod_{k}\prod_{i} q^s({\mathcal{I}}^i_k)}$. Let us assume $ q^s({\mathcal{I}}^i_k)= \delta({{\mathcal{I}}}^i_k-{\breve{{\mathcal{I}}}}^i_k)$ where ${\breve{{\mathcal{I}}}}^i_k$ denotes the point approximation of ${{{\mathcal{I}}}}^i_k$. Consequently, the EM steps are given as 
\subsubsection*{\textnormal{E-Step}}
\begin{align}
	{q^s(\boldsymbol{\mathbf{x}_}{1:K})}&=p(\mathbf{x}_{1:K}|\mathbf{y}_{1:{K}},\breve{\bm{\mathcal{I}}}_{1:K})	\propto	p(\mathbf{x}_{1:K},\breve{\bm{\mathcal{I}}}_{1:K}|\mathbf{y}_{1:{K}})	\label{eqn_sm_E}
\end{align}
\subsubsection*{\textnormal{M-Step}}
\begin{equation}
	{\breve{\mathcal{I}}}^i_k= \underset{{{\mathcal{I}}}^i_k}{\mathrm{argmax}}\big\langle\mathrm{ln}(p(\mathbf{x}_{1:K},{\mathcal{I}}^i_k,\breve{\bm{\mathcal{I}}}^{i-}_k,\breve{\bm{\mathcal{I}}}_{k{-}} |\mathbf{y}_{1:{K}})\big\rangle_{q^s(\mathbf{x}_{1:K})} \label{eqn_sm_M}
\end{equation}
where all ${\hat{\mathcal{I}}}^i_k$ in the M-Step are sequentially updated using the latest estimates. 
\subsubsection{Derivation of $q^s\left(\mathbf{x}_{1:K}\right)$}
With the E-Step in \eqref{eqn_sm_E} we can write 
\begin{align}
	{q^s(\boldsymbol{\mathbf{x}_}{1:K})} &\propto p(\mathbf{y}_{{1:K}}|\mathbf{x}_{{1:K}},\breve{\bm{\mathcal{I}}}_{1:K})p(\mathbf{x}_{{1:K}})\label{eqn_sm_x1}\\	
	{q^s(\boldsymbol{\mathbf{x}_}{1:K})}&\propto \prod_{k} p(\mathbf{y}_{{k}}|\mathbf{x}_{{k}},\breve{\bm{\mathcal{I}}}_{k})p(\mathbf{x}_{{k}}|\mathbf{x}_{{k-1}})\label{eqn_sm_x2}
\end{align}

We can identify that ${q^s(\boldsymbol{\mathbf{x}_}{1:K})}$ can be approximated as a Gaussian distribution from the results of general Gaussian RTS smoothing \cite{sarkka2023bayesian}. Using the forward and backward passes, we can determine the parameters of ${q^s(\boldsymbol{\mathbf{x}_}{k})}\sim\mathcal{N}(\bm{\mathfrak{m}}^{s}_{k},\bm{\mathcal{P}}^{s}_{k})$, which denotes the marginalized densities of ${q^s(\boldsymbol{\mathbf{x}_}{1:K})}$. 
\subsubsection*{Forward pass}
The forward pass essentially involves the filtering equations given as 
\begin{flalign}
	\bm{\mathfrak{m}}^{-}_{k}=&\int \mathbf{f}(\mathbf{x}_{k-1})\  {\mathcal{N}}\left(\mathbf{x}_{k-1}|\bm{\mathfrak{m}}^{+}_{k-1},\bm{\mathcal{P}}^{+}_{k-1}\right)	d\mathbf{x}_{k-1}&\label{eqn_sm_x3a}\\
	\bm{\mathcal{P}}^{-}_{k}=&\int\Big\{( \mathbf{f}(\mathbf{x}_{k-1})-\bm{\mathfrak{m}}^{-}_{k})(\mathbf{f}(\mathbf{x}_{k-1})-\bm{\mathfrak{m}}^{-}_{k})^{\top}&\label{eqn_sm_x3b}\\
	&{\mathcal{N}}({\mathbf{x}_{k-1}}|\bm{\mathfrak{m}}^{+}_{k-1},\bm{\mathcal{P}}^{+}_{k-1})\Big\} d\mathbf{x}_{k-1}+\mathbf{Q}_{k-1}&\label{eqn_sm_x3c}\\
	\bm{\mathfrak{m}}^{+}_k=&\ \bm{\mathfrak{m}}^{-}_k+\bm{\mathcal{K}}_k
	(\mathbf{y}_k-\bm{\nu}_k)	&\label{eqn_sm_x4}\\
	\bm{\mathcal{P}}^{+}_{k}=&\ \bm{\mathcal{P}}^{-}_{k}-\bm{\mathcal{C}}_k\bm{\mathcal{K}}^\top_k &\label{eqn_sm_x5}
\end{flalign}
where
\begin{align}
	\bm{\mathcal{K}}_k=&\ \bm{\mathcal{C}}_k (\bm{\mathcal{U}}_k+{\mathbf{R}}_{k}({\boldsymbol{ \breve{\mathcal{I}} }_{k}}) )^{-1} = \{\bm{\mathcal{C}}_k ( {\mathbf{R}}_{k}^{-1}({\boldsymbol{ \breve{\mathcal{I}} }_{k}})\nonumber\\& {-} {\mathbf{R}}_{k}^{-1}({\boldsymbol{ \breve{\mathcal{I}} }_{k}})(\mathbf{I}+\bm{\mathcal{U}}_k {\mathbf{R}}_{k}^{-1}({\boldsymbol{ \breve{\mathcal{I}} }_{k}}) )^{-1}\bm{\mathcal{U}}_k {\mathbf{R}}_{k}^{-1}({\boldsymbol{ \breve{\mathcal{I}} }_{k}}) )\}\nonumber\\
	\bm{\nu}_k=&\int \mathbf{h}(\mathbf{x}_k)\  {\mathcal{N}}\left(\mathbf{x}_k|\bm{\mathfrak{m}}^{-}_{k},\bm{\mathcal{P}}^{-}_{k}\right)	d\mathbf{x}_k \nonumber\\
	\bm{\mathcal{U}}_k=&\int(\mathbf{h}(\mathbf{x}_k)-\bm{\nu}_k)(\mathbf{h}(\mathbf{x}_k)-\bm{\nu}_k)^{\top}{\mathcal{N}}({\mathbf{x}_k}|\bm{\mathfrak{m}}^{-}_k,\bm{\mathcal{P}}^{-}_{k})d\mathbf{x}_k \nonumber\\
	\bm{\mathcal{C}}_k=&\int(\mathbf{x}_k-\bm{\mathfrak{m}}^{-}_{k})(\mathbf{h}(\mathbf{x}_k)-\bm{\nu}_k)^{\top}{\mathcal{N}}({\mathbf{x}_k}|\bm{\mathfrak{m}}^{-}_{k},\bm{\mathcal{P}}^{-}_{k})d\mathbf{x}_k \nonumber
\end{align}

Note that ${\mathbf{R}}_{k}({\boldsymbol{ \breve{\mathcal{I}} }_{k}})$ and ${\mathbf{R}}_{k}^{-1}({\boldsymbol{ \breve{\mathcal{I}} }_{k}})$ can be evaluated similar to ${\mathbf{R}}_{k}({\boldsymbol{ \hat{\mathcal{I}} }_{k}})$ and ${\mathbf{R}}_{k}^{-1}({\boldsymbol{ \hat{\mathcal{I}} }_{k}})$. 
\subsubsection*{Backward pass}
The backward pass can be completed as
\begin{flalign}
	\bm{\mathcal{L}}_{k+1}  =&\int(\mathbf{x}_k-\bm{\mathfrak{m}}^{+}_k)(\mathbf{f}\left(\mathbf{x}_k\right)-\bm{\mathfrak{m}}_{k+1}^{-})^{\top}{\mathcal{N}}\left(\mathbf{x}_{k}|\bm{\mathfrak{m}}^{+}_{k},\bm{\mathcal{P}}^{+}_{k}\right)d\mathbf{x}_{k}  &\label{eqn_sm_x11}\\
	\bm{\mathcal{G}}_k  = &\ \bm{\mathcal{L}}_{k+1}\left(\bm{\mathcal{P}}_{k+1}^{-}\right)^{-1} \\
	\bm{\mathfrak{m}}_k^{{s}}  = &\ \bm{\mathfrak{m}}^{+}_k+\bm{\mathcal{G}}_k\left(\bm{\mathfrak{m}}_{k+1}^{{s}}-\bm{\mathfrak{m}}_{k+1}^{-}\right) &\label{eqn_sm_m}\\
	\bm{\mathcal{P}}_k^{{s}}  =&\ \bm{\mathcal{P}}^{+}_k+\bm{\mathcal{G}}_k\left(\bm{\mathcal{P}}_{k+1}^{{s}}-\bm{\mathcal{P}}_{k+1}^{-}\right) \bm{\mathcal{G}}_k^{\top} &\label{eqn_sm_P}
\end{flalign}
\subsubsection{Derivation of $\breve{\mathcal{I}}^i_k$}
With the M-Step in \eqref{eqn_sm_M} we can write 
\begin{align}
	{\breve{\mathcal{I}}}^i_k= \underset{{{\mathcal{I}}}^i_k}{\mathrm{argmax}}\big\langle\mathrm{ln}(p(\mathbf{x}_{1:K},{\mathcal{I}}^i_k,\breve{\bm{\mathcal{I}}}^{i-}_k,\breve{\bm{\mathcal{I}}}_{k{-}}|\mathbf{y}_{1:{k}})\big\rangle_{q^s(\mathbf{x}_{1:K})} \label{eqn_sm_i1}
\end{align}
	
Using the Bayes rule we can proceed as
\begin{flalign}
	{\breve{\mathcal{I}}}^i_k=&\ \underset{{{\mathcal{I}}}^i_k}{\mathrm{argmax}}\Big\{ \big\langle\mathrm{ln}(p(\mathbf{y}_{{k}}|\mathbf{x}_{k},{\mathcal{I}}^i_k,\breve{\bm{\mathcal{I}}}^{i-}_k,\stkout{\breve{\bm{\mathcal{I}}}_{k{-}},{\mathbf{y}_{k-} }} ))& \nonumber\\
	&+\mathrm{ln} (p(\mathbf{x}_{1:K},{\mathcal{I}}^i_k,\breve{\bm{\mathcal{I}}}^{i-}_k,\breve{\bm{\mathcal{I}}}_{k{-}}|\mathbf{y}_{k-} )) \big\rangle_{q^s(\mathbf{x}_{1:K})}+const. \Big\} & \label{eqn_sm_i2}
\end{flalign}
which leads to
\begin{flalign}
	{\breve{\mathcal{I}}}^i_k=&\ \underset{{{\mathcal{I}}}^i_k}{\mathrm{argmax}}\Big\{ \big\langle\mathrm{ln}(p(\mathbf{y}_{{k}}|\mathbf{x}_{k},{\mathcal{I}}^i_k,\breve{\bm{\mathcal{I}}}^{i-}_k ))\big\rangle_{q^s(\mathbf{x}_{k})} & \nonumber\\
	&+\mathrm{ln} (p({\mathcal{I}}^i_k|\stkout{{\mathbf{x}_{1:K}},{\breve{\bm{\mathcal{I}}}^{i-}_k},{\hat{\bm{\mathcal{I}}}_{k{-}}},{\mathbf{y}_{k-}}} ))+const.  \Big\} & \label{eqn_sm_i3}
\end{flalign}
which is similar to \eqref{eqn_fl_up2_i} except that the expectation is taken with respect to the marginal smoothing distribution ${q^s(\boldsymbol{\mathbf{x}_}{k})}$. Consequently, $\breve{\mathcal{I}}^i_k$ can be determined as
\begin{align}
	{\breve{\mathcal{I}}}^i_k &= 
	\begin{cases}
		1 & \text{if } \breve{\tau}^i_k \leq 0,\\
		{\epsilon} & \text{if } \breve{\tau}^i_k >0
	\end{cases} \label{Is}
\end{align}
where
\begin{flalign}
	\breve{\tau}^i_k =&\ \Big\{ \mathrm{tr} \big( \bm{\mathcal{W}}_k \triangle{\breve{\mathbf{R}}}_{k}^{-1}  \big) + 
	\ln \Bigg|\textbf{I}-\frac{\mathbf{R}^{-i,i}_k\mathbf{R}^{i,-i}_k (\breve{\mathbf{R}}^{-i,-i}_{k})^{-1}}{{R}^{i,i}_k}\Bigg|& \nonumber\\ &+ \ln(\epsilon)+2\ln\Big(\frac{1}{\theta^i_k}-1  \Big) \Big\} & \label{eqn_sm_up5_i}
\end{flalign}
with
\begin{align}
	\triangle{\breve{\mathbf{R}}}_{k}^{-1}=({\mathbf{R}}_{k}^{-1}( {\mathcal{I}^i_k=1} , \breve{\bm{\mathcal{I}}}^{i-}_k )-{\mathbf{R}}_{k}^{-1}( {\mathcal{I}^i_k=\epsilon} , \breve{\bm{\mathcal{I}}}^{i-}_k ))	
\end{align}
\begin{algorithm}[t!]
	\SetAlgoLined
	Initialize\ $\bm{\mathfrak{m}}^{+}_0,\bm{\mathcal{P}}^{+}_0$;

	\While{not converged}{
		\For{$k=1,2...K$}{
			Initialize $\theta^i_k,{\boldsymbol{ \breve{\mathcal{I}} }_{k}},\mathbf{Q}_k,\mathbf{R}_k$\;	
			Evaluate $\bm{\mathfrak{m}}^{-}_k,\bm{\mathcal{P}}^{-}_k$ with \eqref{eqn_sm_x3a}-\eqref{eqn_sm_x3b}\;
			Evaluate $\bm{\mathfrak{m}}^{+}_k,\bm{\mathcal{P}}^{+}_k$ with \eqref{eqn_sm_x4}-\eqref{eqn_sm_x5}\;
		}
		$\bm{\mathfrak{m}}^{s}_K=\bm{\mathfrak{m}}^{+}_K$\;
		$\bm{\mathcal{P}}^{s}_K=\bm{\mathcal{P}}^{+}_K$ \;
		\For{$k=K-1,...1$}{
			Evaluate $\bm{\mathfrak{m}}^{s}_k,\bm{\mathcal{P}}^{s}_k$ with \eqref{eqn_sm_m}-\eqref{eqn_sm_P}\;
		}	
		\For{$k=1,2...K$}{
			Update ${\breve{\mathcal{I}}^i_k}$ $\forall\ i$ with \eqref{Is}\;
		}
	}
	\caption{The proposed smoother: EMORS}
	\label{Algo2}
\end{algorithm} 
which can be be calculated similar to $\triangle{\hat{\mathbf{R}}}_{k}^{-1}$ in \eqref{eqdelR}. $\breve{\mathbf{R}}^{-i,-i}_{k}$ denotes the submatrix of ${\mathbf{R}}_{k}({\boldsymbol{ \breve{\mathcal{I}} }_{k}})$ corresponding to entries of $\breve{\boldsymbol{\mathcal{I}}}^{i-}_k$ and

\begin{align}
	\bm{\mathcal{W}}_k=\int \left(\mathbf{y}_{k}-\mathbf{h}\left(\mathbf{x}_{k}\right)\right) \left(\mathbf{y}_{k}-\mathbf{h}\left(\mathbf{x}_{k}\right)\right)^{\top} {\mathcal{N}}({\mathbf{x}_k}|\bm{\mathfrak{m}}^{s}_{k},\bm{\mathcal{P}}^{s}_{k})d\mathbf{x}_k\nonumber
\end{align}

The resulting EM-based outlier-robust smoother (EMORS) is outlined as Algorithm \ref{Algo2}. {As EMORS is built on similar lines to EMORF, the same practical considerations hold true including the evaluation of integrals, convergence criterion and selection of parameters.}

\section{Performance Bounds}\label{Bounds}
It is useful to determine the performance bounds of outlier-discarding state estimators considering correlated measurement noise. We evaluate the estimation bounds of filtering and smoothing approaches that are perfect outlier rejectors, having complete knowledge of the instances of outlier occurrences. In particular, we assume that the measurement covariance matrix is a function of perfectly known values of $\bm{\mathcal{I}}_k$ given as $\mathbf{R}_k(\bm{\mathcal{I}}_k)$. In this case, ${\mathcal{I}}^i_k=0$ means rejection of the $i$th corrupted dimension, whereas ${\mathcal{I}}^i_k=1$ denotes inclusion of the $i$th measurement. Resultingly, $\mathbf{R}^{-1}_k(\bm{\mathcal{I}}_k)$ has zeros at the diagonals, rows, and columns corresponding to dimensions for which ${\mathcal{I}}^i_k=0$. Remaining submatrix of $\mathbf{R}^{-1}_k(\bm{\mathcal{I}}_k)$ can be evaluated as the inverse of submatrix of $\mathbf{R}_k$ considering the dimensions with ${\mathcal{I}}^i_k=1$.

Note that we set ${\mathcal{I}}^i_k=\epsilon$, not exactly as $0$, for outlier rejection in the proposed state estimators as it declines inference. However, for evaluating performance bounds this choice is appropriate resulting in perfect outlier rejection. Also note that during robust state estimation, we do not exactly know $\bm{\mathcal{I}}_k$ apriori and model it statistically for subsequent inference. The use of perfectly known $\bm{\mathcal{I}}_k$ for estimation bounds gives us an idea of how well we can estimate the state if outliers are somehow perfectly detected and rejected. 

We evaluate BCRBs for the perfect rejector for the model in \eqref{eqn_model_1}-\eqref{eqn_model_2} that have been corrupted with measurement outliers for both filtering and smoothing. 


\subsection{Filtering}
For the estimation error of $\mathbf{x}_{k}$ during filtering, the BCRB matrix can be written as \cite{van2004detection}

\begin{equation*}
	\text{BCRB}^f_{k}\triangleq ({\mathbf {J}^{+}_{k}})^{-1} \tag{49}
\end{equation*}
where the corresponding filtering Fisher information matrix (FIM) denoted as $\mathbf{J}^{+}_{k}$ can be evaluated recursively as 
\begin{align}
	&\mathbf{J}^{-}_{k}=\mathbf {D}_{k-1}^{22}(1)-\mathbf {D}_{k-1}^{21}\left(\mathbf {J}^{+}_{k-1}+\mathbf {D}_{k-1}^{11}\right)^{-1} \mathbf {D}_{k-1}^{12} \\
	&\mathbf{J}^{+}_{k}=\mathbf{J}^{-}_{k}+\mathbf {D}_{k-1}^{22}(2)
\end{align}
where $\mathbf{J}^{+}_{0}=\langle -\Delta _{\mathbf {x}_{0}}^{\mathbf {x}_{0}} \ln p(\mathbf {x}_{0}) \rangle _{p(\mathbf {x}_{0})}$ and
\begin{align}
	&\Delta _{\Psi }^{\Theta }=\nabla _{\Psi } \nabla _{\Theta }^{\top } 
	\\
	&\nabla _{\Theta }=\left[\frac{\partial }{\partial \Theta _{1}},\ldots, \frac{\partial }{\partial \Theta _{r}}\right]^{\top }
\end{align}

\begin{flalign}
	\mathbf{D}_{k}^{11}&=\langle -\Delta _{\mathbf {x}_{k}}^{\mathbf {x}_{k}} \ln p\left(\mathbf {x}_{k+1} \mid \mathbf {x}_{k}\right)\rangle _{p(\mathbf {x}_{k+1},\mathbf {x}_{k})}
	&\\
	\mathbf {D}_{k}^{12}&=\langle -\Delta _{\mathbf {x}_{k}}^{\mathbf {x}_{k+1}} \ln p\left(\mathbf {x}_{k+1} \mid \mathbf {x}_{k}\right)\rangle _{p(\mathbf {x}_{k+1},\mathbf {x}_{k})}
	&\\
	\mathbf {D}_{k}^{21}&=\langle -\Delta _{\mathbf {x}_{k+1}}^{\mathbf {x}_{k}} \ln p\left(\mathbf {x}_{k+1} \mid \mathbf {x}_{k}\right)\rangle _{p(\mathbf {x}_{k+1},\mathbf {x}_{k})}=\left(\mathbf {D}_{k}^{12}\right)^{\top }
	&\\
	\mathbf {D}_{k}^{22}&= \mathbf {D}_{k}^{22}(1)+\mathbf {D}_{k}^{22}(2)
	&\\
	\mathbf {D}_{k}^{22}(1)&=\langle -\Delta _{\mathbf {x}_{k+1}}^{\mathbf {x}_{k+1}} \ln p\left(\mathbf {x}_{k+1} \mid \mathbf {x}_{k}\right)\rangle _{p(\mathbf {x}_{k+1},\mathbf {x}_{k})} &\\
	\mathbf{D}_{k}^{22}(2)&= \langle -\Delta _{\mathbf {x}_{k+1}}^{\mathbf {x}_{k+1}} \ln p\left(\mathbf {y}_{k+1} \mid \mathbf {x}_{k+1}\right)\rangle _{p(\mathbf {y}_{k+1},\mathbf {x}_{k+1})}& \label{Dk22_2}
\end{flalign}

The bound is valid given the existence of the following derivatives and expectations terms for an asymptotically unbiased estimator \cite{668800}. For the perfect rejector considering the system model in \eqref{eqn_model_1}-\eqref{eqn_model_2} that is infested with observation outliers we can write 
\begin{align}
	\mathbf {D}_{k}^{11}&=\langle \tilde{\mathbf{F}}^{\top}(\mathbf{x}_k) \mathbf {Q}_{k}^{-1} \tilde{\mathbf{F}}(\mathbf{x}_k) \rangle_{p(\mathbf{x}_k)} 
	\label{Dk11} \\
	\mathbf {D}_{k}^{12}&= -\langle \tilde{\mathbf{F}}^{\top}(\mathbf{x}_k) \rangle_{p(\mathbf{x}_k)} \mathbf {Q}_{k}^{-1}
    \label{Dk12}	\\
	\mathbf {D}_{k}^{22}(1)&= \mathbf {Q}_{k}^{-1} \label{Dk22_1} \\
	\mathbf {D}_{k}^{22}(2) &= \langle \tilde{\mathbf {H}}^{\top}(\mathbf{x}_{k+1}) \mathbf {R}_{k+1}^{-1}(\bm{\mathcal{I}}_{k+1}) \tilde{\mathbf {H}}(\mathbf{x}_{k+1}) \rangle _{p(\mathbf {x}_{k+1}) }  \label{Dk22_2_2}
\end{align}
where $\tilde{\mathbf {F}}(.)$ and $\tilde{\mathbf {H}}(.)$ are the Jacobians of the transformations $\mathbf{f}(.)$ and $\mathbf{h}(.)$ respectively. 


\subsection{Smoothing}
Similarly, for the estimation error of $\mathbf{x}_{k}$ during smoothing, the BCRB matrix can be written as \cite{SIMANDL20011703}
\begin{equation}
	\text{BCRB}^s_{k}\triangleq ({\mathbf {J}^{s}_{k}})^{-1} 
\end{equation}
where $\mathbf{J}^s_{K}=\mathbf{J}^{+}_{K}$. We can compute the associated smoothing FIM denoted as $\mathbf{J}^{s}_{k}$ recursively as
\begin{align}
	&\mathbf{J}^{s}_{k}=\mathbf{J}^{+}_{k}+\mathbf {D}_{k}^{11}-\mathbf {D}_{k}^{12}\left(\mathbf {D}_{k}^{22}(1)+\mathbf{J}^s_{k+1}+\mathbf {J}^{-}_{k+1}\right)^{-1} \mathbf {D}_{k}^{21}
\end{align}

\section{Numerical Experiments}\label{exp}
To test the performance of the proposed outlier-resilient state estimators, we carry out several numerical experiments.  We use Matlab on a computer powered by an Intel i7-8550U processor. All the experiments were conducted while considering SI units.

\begin{figure}[h!]
	\centering
	\includegraphics[width=1\linewidth]{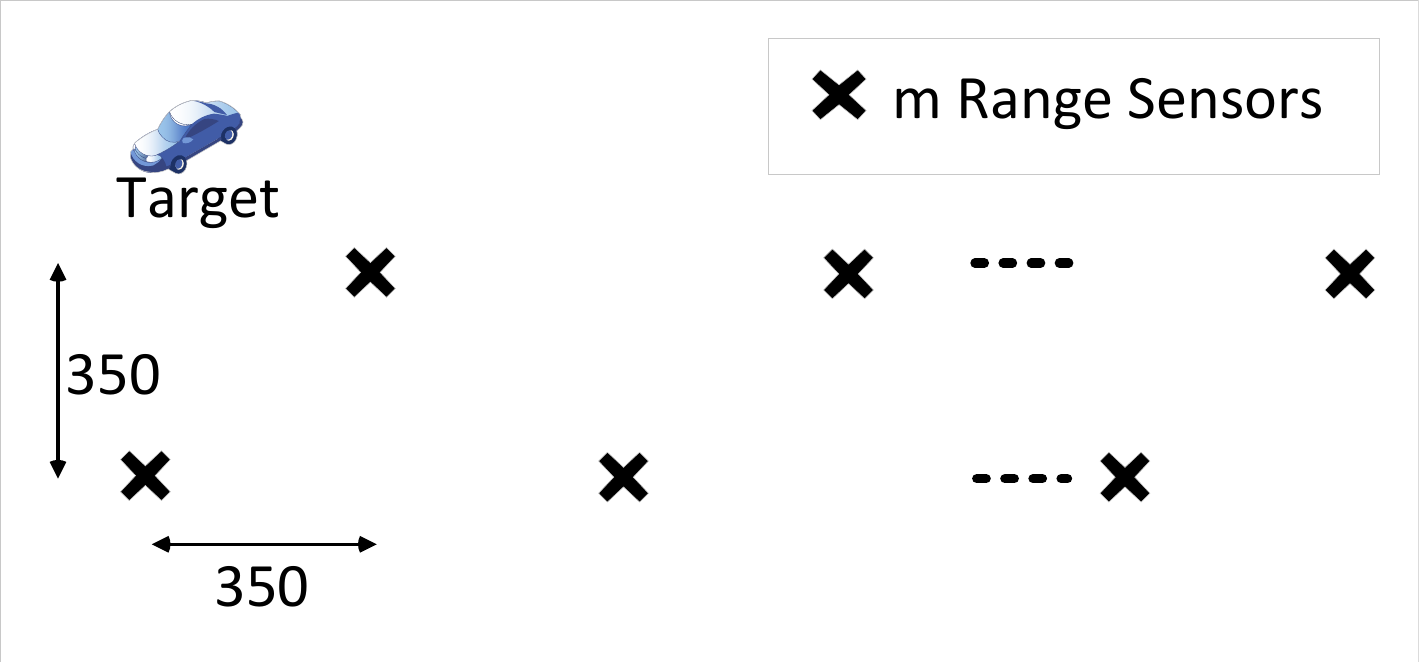}
	\caption{Target tracking test example setup } 
	\label{fig:target}
\end{figure}

For performance evaluation, we resort to a target tracking problem with TDOA-based range measurements inspired by \cite{5977569}. Fig.~\ref{fig:target} shows the setup of the considered example. Owing to the use of a common reference sensor to obtain the TDOA observations, from the difference of the time of arrival (TOA) measurements, the resulting covariance matrix becomes fully populated. 

We consider the process equation for the target assuming an unknown turning rate as \cite{8398426}

\begin{align}
	\mathbf{x}_k=\mathbf{f}(\mathbf{x}_{k-1})+\mathbf{q}_{k-1}\label{eqn_res1a}
\end{align}
with
\begin{align}
	\mathbf{f}(\mathbf{x}_{k-1}) &= \begin{bmatrix} \text{1} & \frac{\text{sin}(\omega_{k-1}{{\zeta}})}{\omega_{k-1}} & \text{0} &  \frac{\text{cos}(\omega_{k-1}{{\zeta}})-\text{1}}{\omega_{k-1}} & \text{0} \\  \text{0} & \text{cos}(\omega_{k-1}{{\zeta}}) &  \text{0} & -\text{sin}(\omega_{k-1}{{\zeta}}) & \text{0}\\  \text{0} & \frac{\text{1}-\text{cos}(\omega_{k-1}{{\zeta}})}{\omega_{k-1}} & \text{1} &\frac{\text{sin}(\omega_{k-1}{{\zeta}})}{\omega_{k-1}} & \text{0} \\  \text{0} &  \text{\text{sin}}(\omega_{k-1}{{\zeta}}) &  \text{0} &  \text{cos}(\omega_{k-1}{{\zeta}}) & \text{0}\\  \text{0} & \text{0} & \text{0} & \text{0} &\text{1} \end{bmatrix} \mathbf{x}_{k-\text{1}}\label{eqn_res1}
\end{align}
where the state vector $\mathbf{x}_k= [a_k,\dot{{a_k}},b_k,\dot{{b_k}},\omega_{k}]^{\top}$
is composed of the 2D position coordinates \(({a_k} , {b_k} )\), the corresponding velocities \((\dot{{a_k}} , \dot{{b_k}} )\), the angular velocity $\omega_{k}$ of the target at time instant $k$, \( {{\zeta}} \) denotes the sampling period, and $\mathbf{q}_{k-\text{1}} \sim N\left(0,\mathbf{Q}_{k-\text{1}}\right)$. $\mathbf{Q}_{k-\text{1}}$ is given in terms of scaling parameters $\eta_1$ and $\eta_2$ as \cite{8398426}
\begin{equation}
	\mathbf{Q}_{k-\text{1}}=\begin{bmatrix} \eta_1 \mathbf{M} & 0 & 0\\0 &\eta_1 \mathbf{M}&0\\0&0&\eta_2
	\end{bmatrix}, \mathbf{M}=\begin{bmatrix} {{\zeta}}^3/3 & {{\zeta}}^2/2\\{{\zeta}}^2/2 &{{\zeta}}
	\end{bmatrix}\nonumber
\end{equation}

Range readings are obtained using $m$ sensors installed in a zig-zag fashion as depicted in Fig.~\ref{fig:target}. The $i$th sensor is located at $\big(a^{\rho_i}=350(i-1),b^{\rho_i}=350\ ((i-1)\mod2)\big)$ for $i=1 \cdots m$. We assume the first sensor as the common sensor for reference resulting in $m-1$ TDOA-based measurements. The nominal measurement equation can be expressed as
\begin{equation}
	\mathbf{y}_k = \mathbf{h}(\mathbf{x}_{k})+\mathbf{r}_k\label{eqn_res2}
\end{equation}
with
\begin{flalign}
	{h^j(\mathbf{x}_k )} =& \Big\{ \sqrt{ (a_{k} - a^{\rho_1})^{2} + (b_{k} - b^{\rho_1})^{2} }  & \nonumber \\ 
	&- \sqrt{ (a_{k} - a^{\rho_{j+1}})^{2} + (b_{k} - b^{\rho_{j+1}})^{2} } \Big\} & \label{eqn_res2b}
\end{flalign}
for $j=1 \cdots m-1$. The corresponding nominal covariance measurement matrix is fully populated given as \cite{5977569}
\begin{align}
	 \mathbf{R}_k= \begin{bmatrix}
		{\sigma^2_1}+{\sigma^2_2}  & \dots &  {\sigma^2_1} \\
		\vdots &  \ddots& \vdots\\
		{\sigma^2_1}  & \cdots &{\sigma^2_1}+{\sigma^2_m}
	\end{bmatrix} \label{sim_R}
\end{align}
where $\sigma^2_i$ is the variance contribution of the $i$th sensor in the resulting covariance matrix. To consider the effect of outliers the measurement equation can be modified as
\begin{equation}
	\mathbf{y}_k = \mathbf{h}(\mathbf{x}_{k})+\mathbf{r}_k+\mathbf{o}_k\label{sim_o}
\end{equation}
where $\mathbf{o}_k$ produces the effect of outliers in the measurements and is assumed to obey the following distribution 
\begin{align}
	p(\mathbf{o}_k)&=\prod_{j=1}^{m-1} \mathcal{J}^j_k {\mathcal{N}}(o^{j}_k|0,\gamma ({\sigma^2_1}+{\sigma^2_j})) \label{eq_sim_o}
\end{align}
where $\mathcal{J}^j_k$ is a Bernoulli random variable, with values $0$ and $1$, that controls whether an outlier in the $j$th dimension occurs. Let $\lambda$ denote the probability that a sensor's TOA measurement is affected. Therefore, the probability that no outlier appears in the $j$th dimension, corresponding to $\mathcal{J}^j_k=0$, is $(1-\lambda)^2$ since the first sensor is a common reference for the TDOA-based measurements. We assume that the TOA measurements are independently affected and the corruption of the first TOA observation affects all the measurements. Similarly, the parameter $\gamma$ controls the variance of an outlier in each dimension respectively. Using the proposed model we generate the effect of outliers in the data. 

For filtering performance comparisons, we choose a hypothetical Gaussian filter that is a perfect rejector having apriori knowledge of all outlier instances. We also consider the generalized and independent VBKF estimators \cite{9286419}, referred from hereon as {Gen. VBKFs} and Ind. VBKF, for comparisons. In Gen. VBKFs, we set the design parameter as $N=1$ and {$N=10$ as originally reported}. Lastly, we use the derived BCRB-based filtering lower bounds to benchmark the performance of all the filters. Similarly, for smoothing we use the counterparts of all the considered filters i.e. a perfect outlier-rejecting general Gaussian RTS smoother and the generalized/independent VBKF-based RTS smoothers denoted as {Gen. VBKSs} and Ind. VBKS.

For simulations the following values of parameters are used: the initial state $\mathbf{x}_0= [0,1,0,-1,-0.0524] ^\text{T}$, $\zeta=1$, $\eta_1=0.1$, $\eta_2=1.75\times10^{-4}$, ${\sigma^2_j}=10$. The initialization parameters of estimators are: {${\mathbf{m}}^+_{0} \sim \mathcal{N}(\mathbf{x}_0,\mathbf{P}^+_{0}$), $\mathbf{P}^+_{0}=\mathbf{Q}_{k}$}, $\epsilon=10^{-6}$ and $\theta^i_k=0.5~\forall~i$. For {fairness we use UT in each method} for approximating the Gaussian integrals \cite{wan2001unscented}, in all the considered methods. Resultingly, the Unscented Kalman Filter (UKF) becomes the core inferential engine for all the techniques.  

In each method, UT parameters are set as $\alpha=1$, $\beta=2$, and $\kappa=0$. Moreover, we use the same threshold of $10^{-4}$ for the convergence criterion in each algorithm. Other parameters for VBKFs/VBKSs are assigned values as originally documented. All the simulations are repeated with a total time duration {{$K=100$}} and $100$ independent MC runs. Moreover, we use box and whisker plots to visualize all the results.  


\subsection{Filtering Performance}
We assess the relative filtering performance under different scenarios. 
\begin{figure}[h!]
	\centering
	\includegraphics[width=\linewidth]{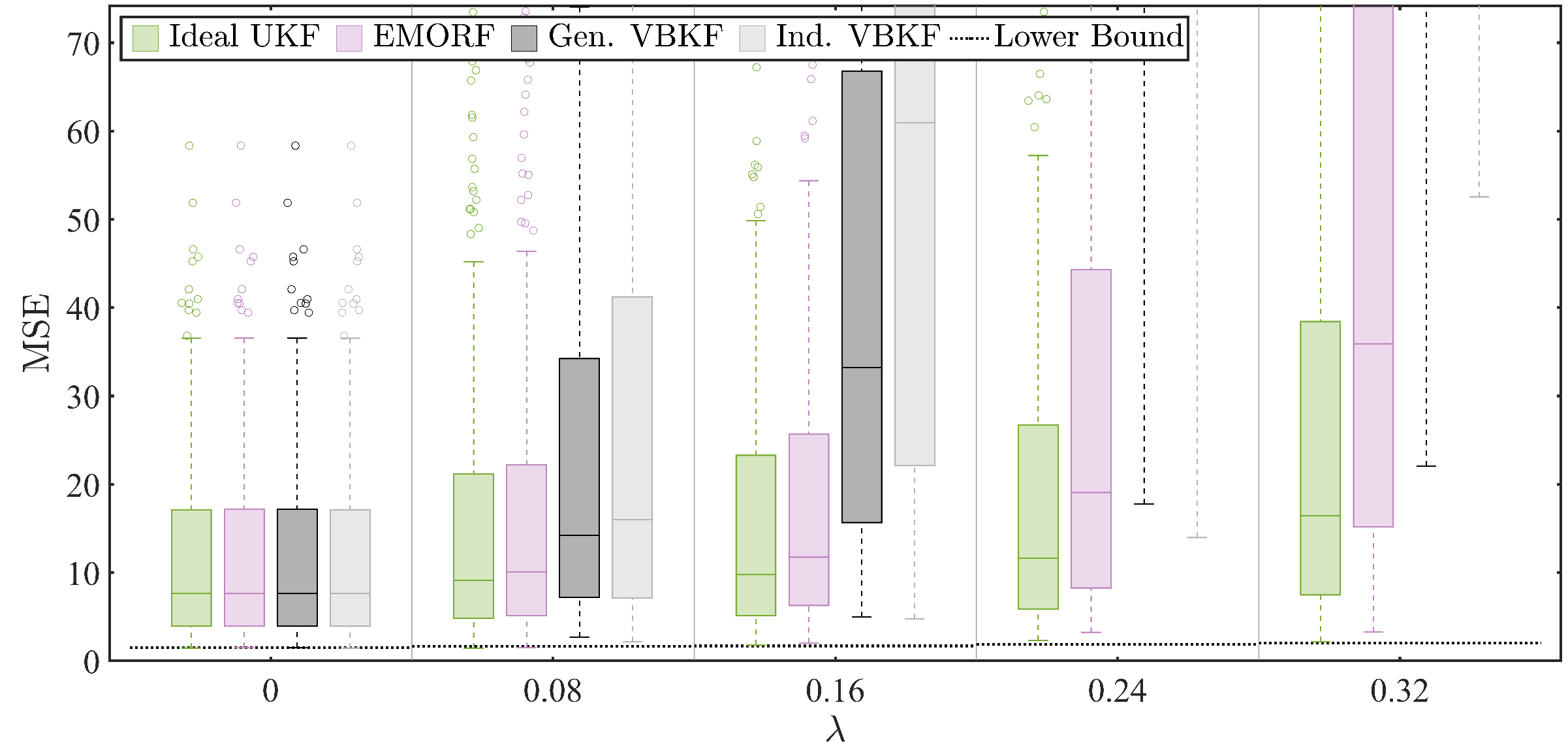}
	\caption{MSE vs $\lambda$ $(m=10,\gamma=1000)$} 
	\label{fig:fil_lambda}
\end{figure}

First, we choose $10$ number of sensors with $\gamma=1000$ and increase the TOA contamination probability $\lambda$. Fig.~\ref{fig:fil_lambda} shows the mean squared error (MSE) of the state estimate of each filter as $\lambda$ is increased. For $\lambda=0$ all the filters essentially work as the standard UKF having similar performance. As $\lambda$ increases, MSE of each method and the lower bound value are seen to increase. The hypothetical ideal UKF exhibits the best performance followed by the proposed EMORF, Gen. VBKFs, and Ind. VBKF respectively. {Gen. VBKF ($N=10$) generally tends to exhibit slightly improved performance as compared to Gen. VBKF ($N=1$) especially at higher values of $\lambda$}. The trend remains the same for each $\lambda$. Similar patterns have been observed for other combinations of $m$ and $\gamma$. Performance degradation of Ind. VBKF as compared to EMORF and Gen. VBKFs is expectable as it ignores the measurement correlations during filtering. We find EMORF to be generally more robust in comparison to Gen. VBKFs. Our results are not surprising given that we found the modified selective observation rejecting (mSOR)-UKF to be more resilient to outliers as compared to the modified outlier-detecting (mOD)-UKF \cite{chughtai2022outlier}, which are designed for independent measurements having similar structures to EMORF and VBKF respectively. 

\begin{figure}[h!]
	\centering
	\includegraphics[width=\linewidth]{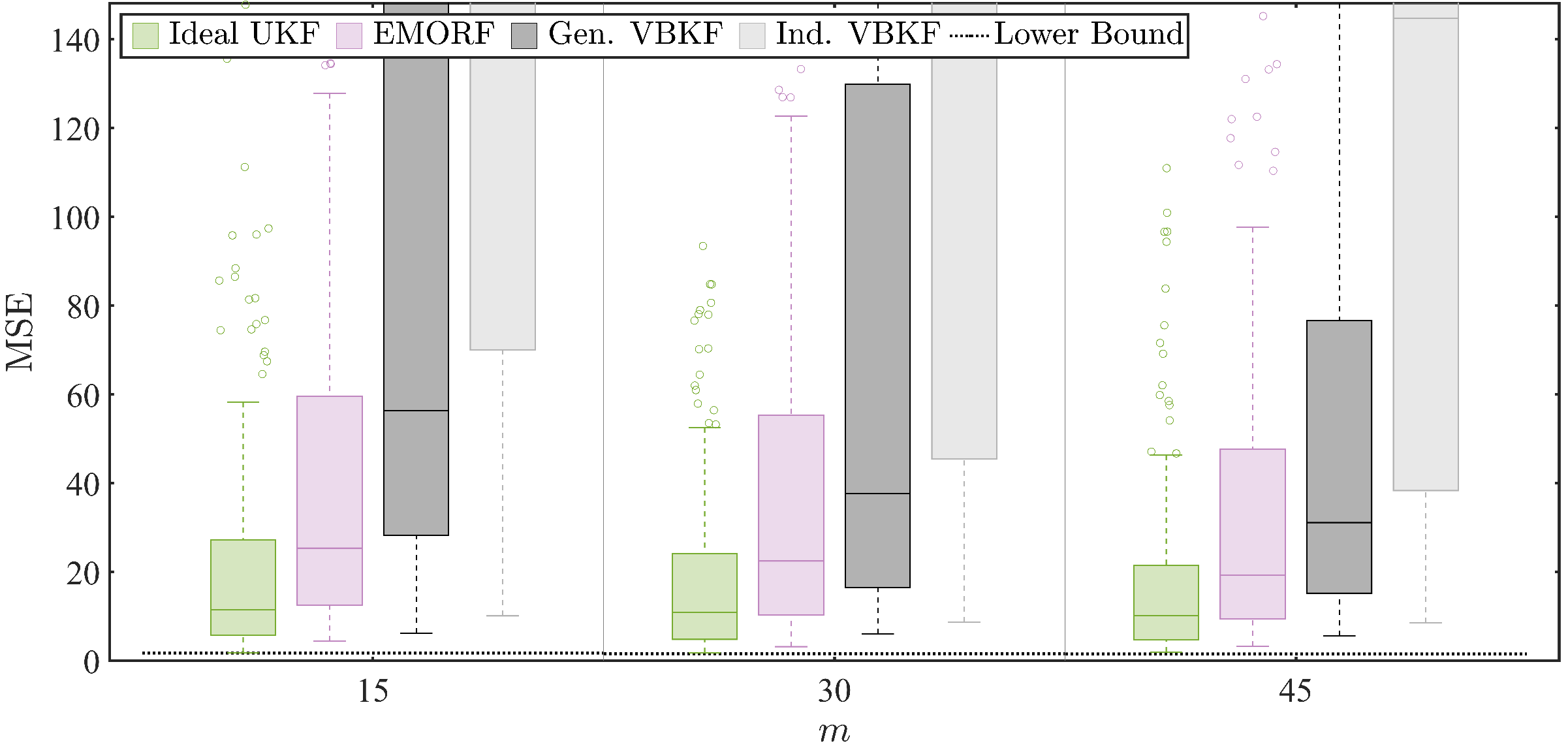}
	\caption{MSE vs $m$ ($\lambda=0.3$ and $\gamma=200$) } 
	\label{fig:fil_sensors}
\end{figure}
Next, we vary the number of sensors and assess the estimation performance of the filters. Fig.~\ref{fig:fil_sensors} shows the MSE of each method as the number of sensors is increased with {$\lambda=0.3$} and {$\gamma=200$}. As expected, the error bound and MSE of each filter decrease with increasing number of sensors since more sources of information become available. We see a pattern similar to the previous case with the best performance exhibited by the hypothetical ideal UKF followed by EMORF, Gen. VBKFs, and Ind. VBKF respectively. {In this simulation setting also, we find Gen. VBKF ($N=10$) to generally have marginally  improved performance as compared to Gen. VBKF ($N=1$).} Moreover, we have observed similar trends for other values of $\lambda$ and $\gamma$ as well.
\begin{figure}[h!]
	\centering
	\includegraphics[width=\linewidth]{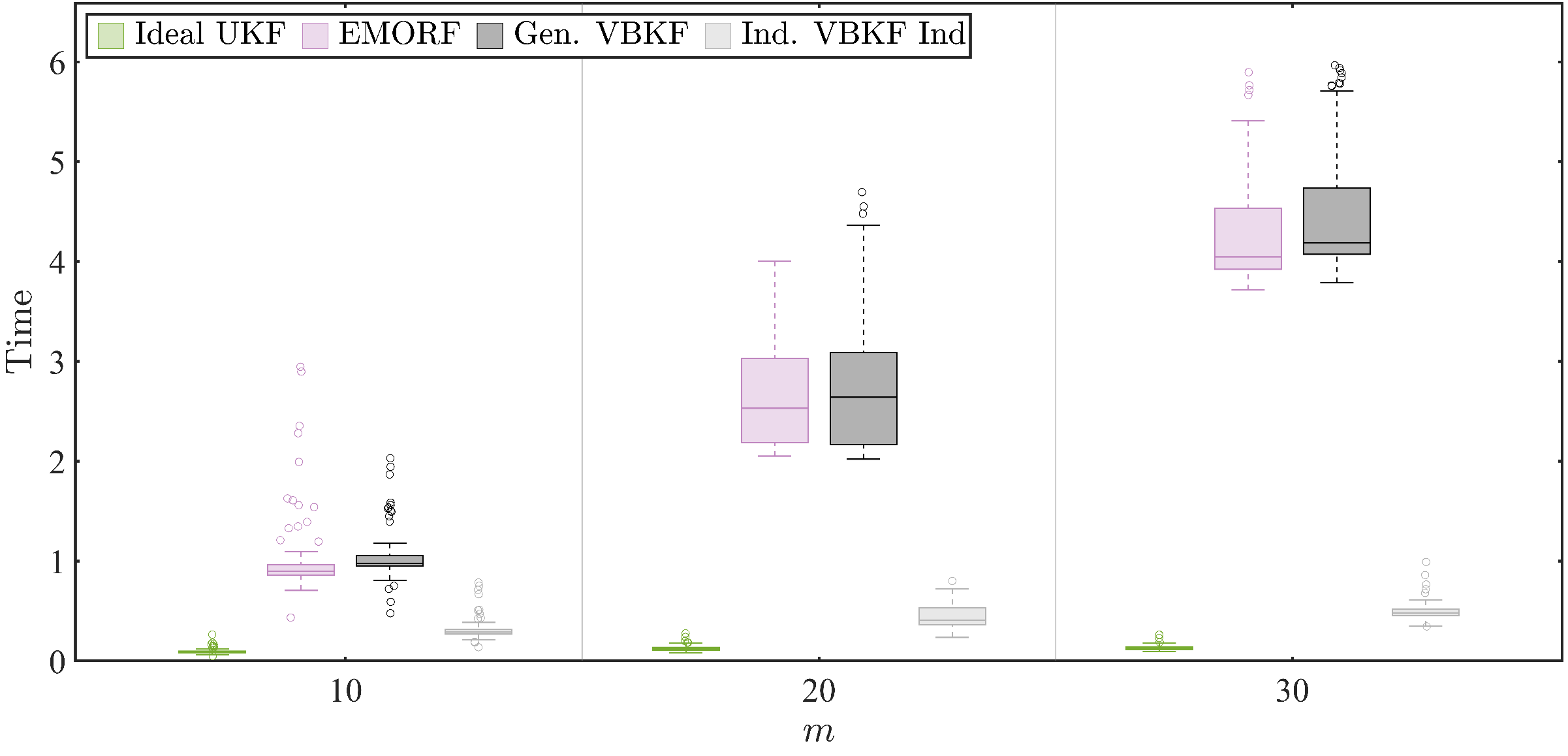}
	\caption{Computational Time vs $m$ ($\lambda=0.3$ and $\gamma=200$) } 
	\label{fig:fil_times}
\end{figure} 

Subsequently, we evaluate the processing overhead of each algorithm by varying the number of sensors. Fig.~\ref{fig:fil_times} shows the execution time taken by each algorithm as the number of sensors is increased with {$\lambda=0.3$} and {$\gamma=200$}. We observe that the ideal UKF and Ind. VBKF take lesser time for execution having a complexity of $\mathcal{O}(m^3)$. However, EMORF and Gen. VBKFs induce more computational overhead, having a complexity of $\mathcal{O}(m^4)$, due to utilization of matrix inverses and determinants for evaluating each of the $\mathcal{I}^i_k$ and $\mathbf{z}^{(i)}_t$ $\forall\ i=1\cdots m$ in EMORF and Gen. VBKFs respectively. This is the cost we pay for achieving robustness with correlated measurement noise. {Gen. VBKF ($N=10$) results in much higher computational cost as compared to Gen. VBKF ($N=1$) due evaluation of complicated expectation expressions. The gains in terms of error reduction do not appear to justify the increase in computational cost in this setting.} We find that EMORF generally takes less processing time as compared to {Gen. VBKF $(N=1)$} as shown in Fig.~\ref{fig:fil_times}. Moreover, similar performance has been observed for other combinations of $\lambda$ and $\gamma$. This can be attributed to a simpler model being employed in EMORF resulting in reduced computations. 

\subsection{Smoothing Performance}
For smoothing we perform analogous experiments and observe similar performance.  
\begin{figure}[h!]
	\centering
	\includegraphics[width=\linewidth]{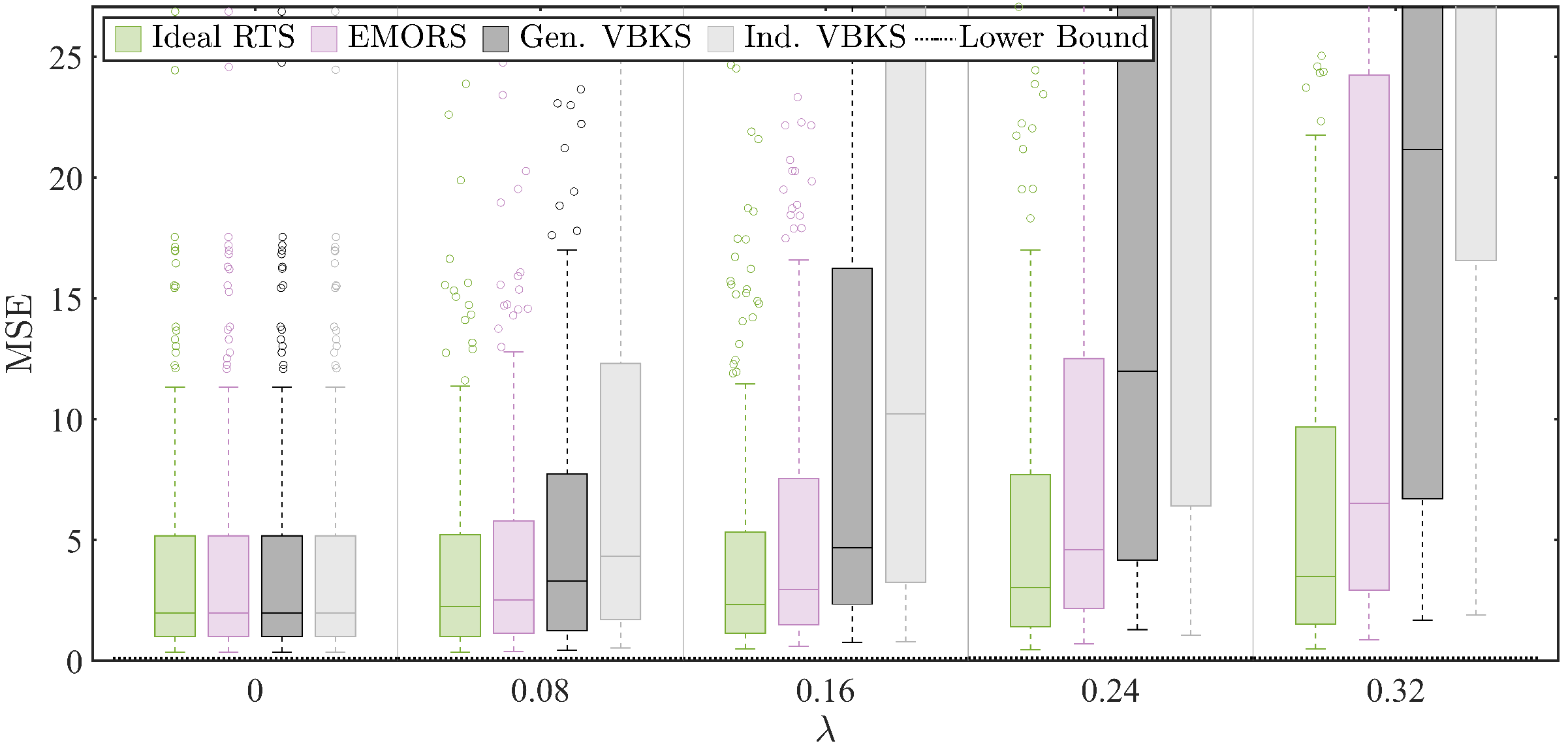}
	\caption{MSE vs $\lambda$ $(m=10,\gamma=500)$} 
	\label{fig:lam_smo}
\end{figure}

First, we choose {$10$} number of sensors and increase the TOA contamination probability $\lambda$ with {$\gamma=500$}. Fig.~\ref{fig:lam_smo} shows how MSE of the state estimate of each smoother changes as $\lambda$ is increased. Similar to filtering, we observe that MSE of each estimator grows with increasing $\lambda$ including the BCRB-based smoothing lower bound. The hypothetical RTS smoother performs the best followed by the proposed EMORS, {Gen. VBKS ($N=10$), Gen. VBKS ($N=1$)} and Ind. VBKS respectively. The trend remains the same for each $\lambda$. Similar patterns have been seen for other combinations of $m$ and $\gamma$. 

\begin{figure}[h!]
	\centering
	\includegraphics[width=\linewidth]{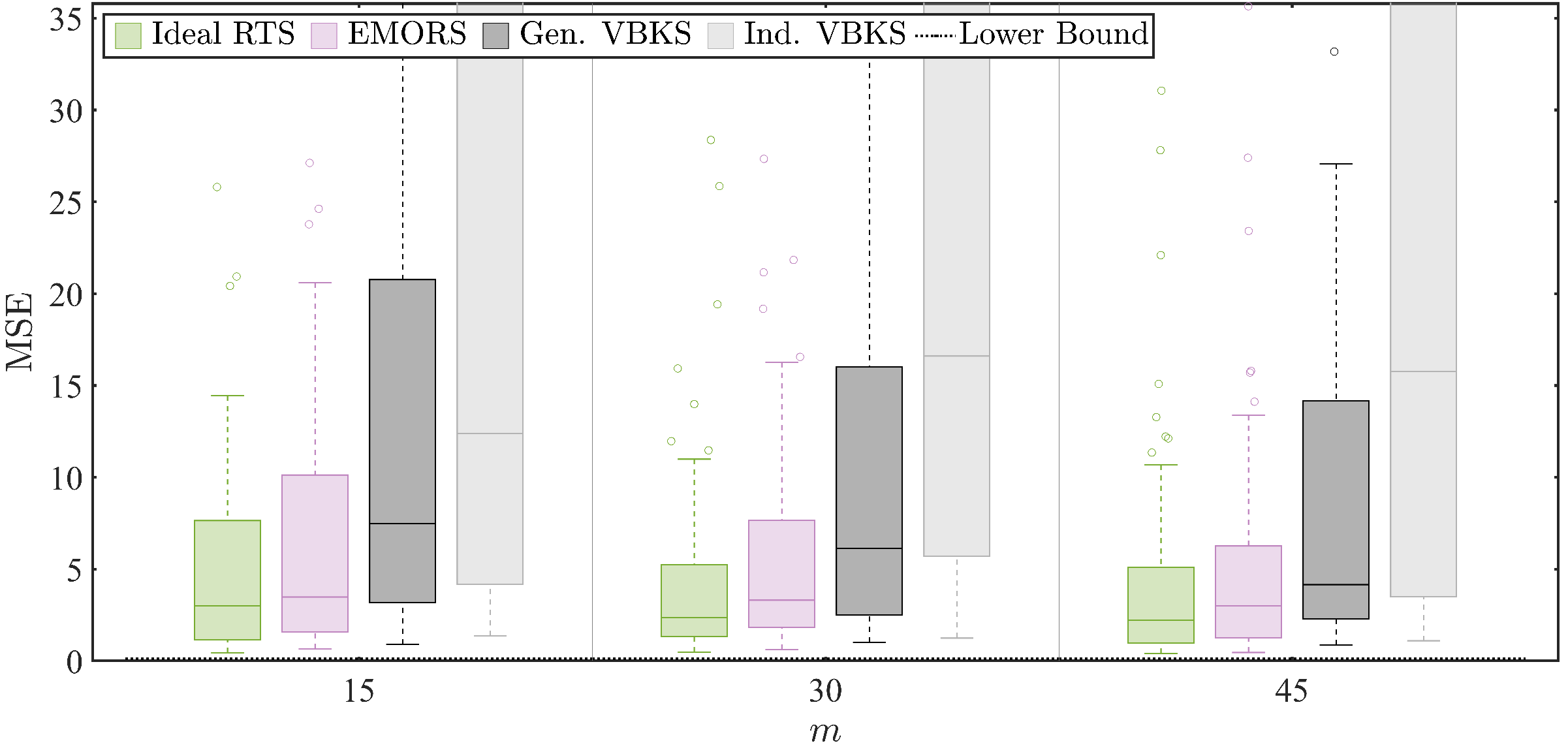}
	\caption{MSE vs $m$ ($\lambda=0.4$ and $\gamma=100$) } 
	\label{fig:smo_sensors}
\end{figure}
Thereafter, we assess the estimation performance of the filters by varying the number of sensors. Fig.~\ref{fig:smo_sensors} depicts MSE of each estimator as the number of sensors increase with {$\lambda=0.4$} and {$\gamma=100$}. MSE for each smoother decreases with growing number of sensors including the BRCB-based lower bound. The hypothetical RTS smoother is {generally} the best performing followed by EMORS, {Gen. VBKS ($N=10$), Gen. VBKS ($N=1$)}, and Ind. VBKS respectively. We have observed similar trends for other values of $\lambda$ and $\gamma$ as well.

\begin{figure}[h!]
	\centering
	\includegraphics[width=\linewidth]{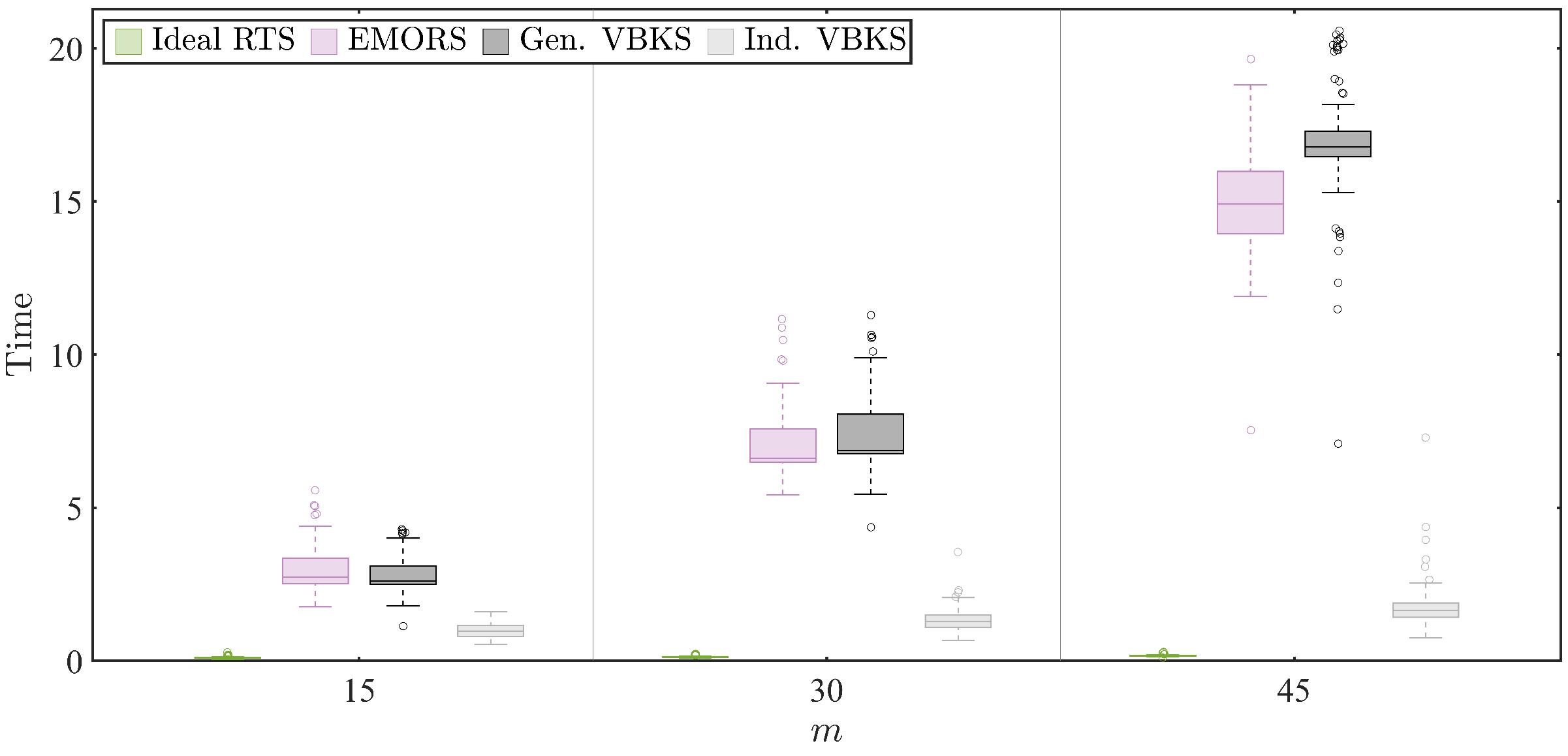}
	\caption{Computational Time vs $m$ ($\lambda=0.4$ and $\gamma=100$) } 
	\label{fig:smo_times}
\end{figure}
Lastly, we evaluate the computational overhead of each algorithm by varying the number of sensors. Fig.~\ref{fig:smo_times} shows the time each method takes as the number of sensors is increased with {$\lambda=0.4$} and {$\gamma=100$}. We observe similar patterns as for filtering. The ideal RTS smoother and Ind. VBKS having a complexity of $\mathcal{O}(m^3)$ take lesser execution time. EMORS and Gen. VBKSs having a complexity $\mathcal{O}(m^4)$ are more time consuming {with Gen. VBKS $(N=10)$ taking much higher computational time than Gen. VBKS $(N=1)$}.  EMORS generally induces lesser computing overhead as compared to {Gen. VBKS ($N=1$)} as depicted in Fig.~\ref{fig:smo_times}. We have observed similar patterns for different combinations of $\lambda$ and $\gamma$.

{\subsection*{Other simulations} 
We repeated the numerical experiments for all the considered filtering and smoothing methods under other scenarios as well. In particular, we replaced the Gaussian distribution in \eqref{eq_sim_o} to Uniform and Laplace distributions and experimented with different distributional parameters. We generally find a similar trend of comparative performance as depicted in Fig.~\ref{fig:fil_lambda}-Fig.~\ref{fig:smo_times}. We also tested the performance of EMORF/S with other choice of the prior parameter $\theta_k^i$. Following the lines of SORF, we repeated the simulations with the prior parameter $\theta_k^i$ drawn from a Uniform density defined over a range $0.05$ to $0.95$ and got similar results as above. These results indicate that the performance of EMORF/S is robust to the changes in the prior parameter. However, the default value of $\theta_k^i=0.5$ is preferable unless some specific prior information regarding the rate of outliers is available (see Appendix \ref{FifthAppendix} for more discussion).}

\section{Conclusion}\label{conc}
We consider the problem of outlier-robust state estimation assuming the existence of measurement noise correlation. Given their advantages, resorting to tuning-free learning-based approaches is an attractive option in this regard. Identifying the shortcomings of such existing VB-based tractable methods, we propose EMORF and EMORS. Since the standard VB approach entails significant processing complexity, we adopt EM in our algorithmic constructions. We can conclude that the presented methods are simpler and hence more practicable as compared to the state-of-the-art Gen. {VBKFs/VBKSs}, devised for the same conditions. This is possible due to the reduction of inference parameters resulting from the proposal of an uncomplicated model. Also, the need of the specialized digamma function during implementation is obviated. In addition, numerical experiments in an illustrative TDOA-based target tracking example suggest further merits of the proposed methods. We find that EMORF/EMORS generally exhibit lesser errors as compared to Gen. and Ind. {VBKFs/VBKSs} in different scenarios of the example. Moreover, though the complexity order of EMORF/EMORS and Gen. {VBKFs/VBKSs} is the same, the proposed estimators are found to be computationally more competitive in general for different test conditions. These merits make the proposed state estimators worthy candidates for implementation in relevant scenarios.   

\appendices

\section{Evaluating ${\mathbf{R}}_{k}^{-1}({\boldsymbol{ \hat{\mathcal{I}} }_{k}})$}\label{FirstAppendix}
For evaluating ${\mathbf{R}}_{k}^{-1}({\boldsymbol{ \hat{\mathcal{I}} }_{k}})$ we consider that ${\mathbf{R}}_{k}({\boldsymbol{ \hat{\mathcal{I}} }_{k}})$ can easily be rearranged by swapping rows/columns depending on $\hat{\boldsymbol{\mathcal{I}}}_k$ as  
\begin{align}
	{{\bm{\mathfrak{R}}}}_{k}({\boldsymbol{ \hat{\mathcal{I}} }_{k}})= 
	\begin{bmatrix}
		\bm{\mathtt{R}}_k / {\epsilon}  & \mathbf{0}  \\
		\mathbf{0} &  \hat{\mathbf{R}}_{k}
	\end{bmatrix}  
\end{align}

where $\bm{\mathtt{R}}_k$ is a sub-matrix with diagonal entries of  ${\mathbf{R}}_k$.  $\hat{\mathbf{R}}_{k}$ contains the rest of the fully populated submatrix of $	{\mathbf{R}}_{k}({\boldsymbol{ \hat{\mathcal{I}} }_{k}})$ corresponding to entries of $\hat{\boldsymbol{\mathcal{I}}}_k=1$. Inversion of ${\bm{\mathfrak{R}}}_{k} ({\boldsymbol{ \hat{\mathcal{I}} }_{k}})$ results in
\begin{align}
	{\bm{\mathfrak{R}}}_{k}^{-1}({\boldsymbol{ \hat{\mathcal{I}} }_{k}})=& \begin{bmatrix}
		{\epsilon} \bm{\mathtt{R}}^{-1}_k & \mathbf{0}  \\
		\mathbf{0} &  \hat{\mathbf{R}}^{-1}_{k}
	\end{bmatrix} \label{eqn_fl_up4_x}
\end{align}

Finally, ${\bm{\mathfrak{R}}}_{k}^{-1}({\boldsymbol{ \hat{\mathcal{I}} }_{k}})$ can be swapped accordingly to obtain the required matrix ${\mathbf{R}}_{k}^{-1}({\boldsymbol{ \hat{\mathcal{I}} }_{k}})$ . 

%
\section{Simplifying $\hat{\tau}^i_k$
}\label{SecondAppendix}
We can swap the \textit{i}th row/column entries of  ${\mathbf{R}}_{k}( {\mathcal{I}^i_k} =1 , \hat{\bm{\mathcal{I}}}^{i-}_k )$ and ${\mathbf{R}}_{k}( {\mathcal{I}^i_k} =1 , \hat{\bm{\mathcal{I}}}^{i-}_k )$ with the first row/column elements to obtain  
\begin{align}
	|{\bm{\mathfrak{R}}}_{k}( {\mathcal{I}^i_k} =1 , \hat{\bm{\mathcal{I}}}^{i-}_k )|&= 
	\begin{vmatrix} 
		R^{i,i}_k & \mathbf{R}^{i,-i}_k \\
		\mathbf{R}^{-i,i}_k & \hat{\mathbf{R}}^{-i,-i}_{k}
	\end{vmatrix} \label{eqn_fl_up6_i} \\
	|{\mathbf{R}}_{k}( {\mathcal{I}^i_k} =\epsilon , \hat{\bm{\mathcal{I}}}^{i-}_k )|&= 
	\begin{vmatrix}
		R^{i,i}_k/\epsilon & \mathbf{0} \\
		\mathbf{0} & \hat{\mathbf{R}}^{-i,-i}_{k}
	\end{vmatrix}  \label{eqn_fl_up7_i}
\end{align}

Consequently, we can write
\begin{align}
	&\ln\Big(\frac{|{\mathbf{R}}_{k}( {\mathcal{I}^i_k=1} , \hat{\bm{\mathcal{I}}}^{i-}_k ) |}{|{\mathbf{R}}_{k}( {\mathcal{I}^i_k=\epsilon} , \hat{\bm{\mathcal{I}}}^{i-}_k )|}\Big)=\ln\Big(\frac{|{\bm{\mathfrak{R}}}_{k}( {\mathcal{I}^i_k=1} , \hat{\bm{\mathcal{I}}}^{i-}_k ) |}{|{\bm{\mathfrak{R}}}_{k}( {\mathcal{I}^i_k=\epsilon} , \hat{\bm{\mathcal{I}}}^{i-}_k )|}\Big)\nonumber \\ 
	&= \ln \Bigg|\textbf{I}-\frac{\mathbf{R}^{-i,i}_k\mathbf{R}^{i,-i}_k (\hat{\mathbf{R}}^{-i,-i}_{k})^{-1}}{{R}^{i,i}_k}\Bigg|+ \ln(\epsilon) 
\end{align}
where we have used the following property from matrix algebra \cite{zhang2006schur}
\begin{align}
	\begin{vmatrix}
		\bm{\mathsf{A}} & \bm{\mathsf{B}} \\
		\bm{\mathsf{C}} & \bm{\mathsf{D}}
	\end{vmatrix}= |\bm{\mathsf{A}}| |\bm{\mathsf{D}}-\bm{\mathsf{C}}\bm{\mathsf{A}}^{-1}\bm{\mathsf{B}}|  \nonumber
\end{align}

Resultingly, we can simplify \eqref{eqn_fl_up5_i} to \eqref{eqn_fl_up10_i}.

\section{Evaluating $\triangle{\hat{\mathbf{R}}}_{k}^{-1}$} \label{ThirdAppendix}
To avoid redundant calculations during the evaluation of $\triangle{\hat{\mathbf{R}}}_{k}^{-1}$, we can first swap the \textit{i}th row/column elements of matrices with the first row/column entries in \eqref{eqdelR} to obtain
\begin{align}
	\triangle{\hat{\bm{\mathfrak{R}}}}_{k}^{-1}&=({\bm{\mathfrak{R}}}_{k}^{-1}( {\mathcal{I}^i_k=1} , \hat{\bm{\mathcal{I}}}^{i-}_k )-{\bm{\mathfrak{R}}}_{k}^{-1}( {\mathcal{I}^i_k=\epsilon} , \hat{\bm{\mathcal{I}}}^{i-}_k ))\nonumber\\
	&=\begin{bmatrix} 
		R^{i,i}_k & \mathbf{R}^{i,-i}_k \\
		\mathbf{R}^{-i,i}_k & \hat{\mathbf{R}}^{-i,-i}_{k}
	\end{bmatrix}^{-1}-\begin{bmatrix} 
		R^{i,i}_k/\epsilon & \mathbf{0}\\
		\mathbf{0} & \hat{\mathbf{R}}^{-i,-i}_{k}
	\end{bmatrix}^{-1}\label{app3_delr}
\end{align}

To simplify \eqref{app3_delr}, we use the following property from matrix algebra \cite{zhang2006schur} 
\begin{align}
	\begin{bmatrix}
		\bm{\mathsf{A}} & \bm{\mathsf{B}} \\
		\bm{\mathsf{C}} & \bm{\mathsf{D}}
	\end{bmatrix}^{-1}= 	
	\begin{bmatrix}
		\bm{\mathsf{S}}^{-1} & -\bm{\mathsf{S}^{-1} \bm{\mathsf{B}} \bm{\mathsf{D}}^{-1} } \\
		-\bm{\mathsf{D}}^{-1} \bm{\mathsf{C}} \bm{\mathsf{S}^{-1} } & \bm{\mathsf{D}}^{-1}+\bm{\mathsf{D}}^{-1} \bm{\mathsf{C}} \bm{\mathsf{S}}^{-1} \bm{\mathsf{B}} \bm{\mathsf{D}}^{-1}  
	\end{bmatrix}  \nonumber
\end{align}
where $\bm{\mathsf{S}}$ is the Schur's complement of $\bm{\mathsf{D}}$ given as $\bm{\mathsf{S}}=\bm{\mathsf{A}}-\bm{\mathsf{B}}\bm{\mathsf{D}}^{-1}\bm{\mathsf{C}}$. As a result, we obtain
\begin{align}
	\triangle{\hat{\bm{\mathfrak{R}}}}_{k}^{-1}&=\begin{bmatrix}
		\bm{\Xi}^{i,i} & \bm{\Xi}^{i,-i} \\
		\bm{\Xi}^{-i,i} & \bm{\Xi}^{-i,-i}  \label{del_t}
	\end{bmatrix}
\end{align}
where the expressions of $\bm{\Xi}^{i,i},\bm{\Xi}^{i,-i},\bm{\Xi}^{-i,i}$ and $\bm{\Xi}^{-i,-i}$ are given in \eqref{xi1}-\eqref{xi4}. The redundant calculations in \eqref{xi1}-\eqref{xi4} can be computed once and stored for further computations e.g. $({R^{i,i}_k-\mathbf{R}^{i,-i}_k(\hat{\mathbf{R}}^{-i,-i}_{k})^{-1} \mathbf{R}^{-i,i}_k})^{-1}$ and $(\hat{\mathbf{R}}^{-i,-i}_{k})^{-1}$. Lastly, the first row/column entries of $\triangle{\hat{\bm{\mathfrak{R}}}}_{k}^{-1}$ are interchanged to the actual $i$th row/column positions to obtain the required  $\triangle{\hat{\mathbf{R}}}_{k}^{-1}$.

{
	\section{Factorization of $p(\mathbf{x}_k,\bm{\mathcal{I}}_k|\mathbf{y}_{1:{k}})$ and the associated computational overhead} \label{ZerothAppendix}
	Note that for better accuracy we can factorize $p(\mathbf{x}_k,\bm{\mathcal{I}}_k|\mathbf{y}_{1:{k}})\approx q^f(\mathbf{x}_k) q^f(\bm{\mathcal{I}}_k)$ instead of forcing independence between all ${\mathcal{I}}^i_k$. In this case, expression for evaluating  $q^f(\mathbf{x}_k)$ remains same as in \eqref{eqn_vb_1}. However, this choice leads to the following VB marginal distribution of $\bm{\mathcal{I}}_k$
	\begin{equation}
		q^f(\bm{\mathcal{I}}_k)\propto \exp \big( \big\langle \mathrm{ln} ( p(\mathbf{x}_k,\bm{\mathcal{I}}_k|\mathbf{y}_{1:k}))\big\rangle_{q^f(\mathbf{x}_k)  }\ \big) \label{eqn_vb_I}
	\end{equation}
	
	This results in a modified M-Step in the EM algorithm given as
	\subsubsection{\textnormal{M-Step}}
	\begin{equation}
		{ \hat {\bm{\mathcal{I}}} }_k= \underset{\bm{{\mathcal{I}}}_k}{\mathrm{argmax}}\big\langle\mathrm{ln}(p(\mathbf{x}_k,{\bm{\mathcal{I}}}_k|\mathbf{y}_{1:{k}}))\big\rangle_{q^f(\mathbf{x}_k)} \label{eqn_fl_M2}
	\end{equation} 
	
	Proceeding further, with the prediction and update steps during inference, we can arrive at 
	\begin{flalign}
		{\hat{\bm{\mathcal{I}}}}_k=& \ \underset{{\bm{\mathcal{I}}}_k}{\mathrm{argmax}} \Big\{ {-}\mfrac{1}{2} \mathrm{tr} \big( \mathbf{W}_k {\mathbf{R}}_{k}^{-1}(  {\bm{\mathcal{I}}}_k )     \big) {-}\mfrac{1}{2}	\ln|{\mathbf{R}}_{k}( {\bm{\mathcal{I}}}_k ) |& \nonumber \\
		&+\ln\big( (1-\theta_{k}^{i})  \delta(\mathcal{I}_{k}^{i}-\epsilon)+\theta_{k}^{i} \delta ({\mathcal{I}}_{k}^{i}-1) \big) \Big\} & \label{eqn_fl_I2}
	\end{flalign}
	
	It is not hard to notice that determining ${\hat{\bm{\mathcal{I}}}}_k$ using \eqref{eqn_fl_I2} involves tedious calculations. In fact, we run into the same computational difficulty that we have been dodging. To arrive at the result, we need to evaluate the inverses and determinants, for each of the $2^m$ combinations corresponding to the entries of ${\bm{\mathcal{I}}}_k$. This entails the bothering complexity level of $\mathcal{O}(m^3 2^m)$. 
	
}

{
\section{Connection between EMORF and SORF \cite{chughtai2022outlier}}\label{ForthAppendix}
Since the construction of EMORF is motivated by SORF, it is insightful to discuss their connection. We derived SORF considering a diagonal measurement covariance matrix $\mathbf{R}_k$ \cite{chughtai2022outlier}. In SORF, we used distributional estimates for $\bm{\mathcal{I}}_k$ since it did not induce any significant computational strain. It is instructive to remark here that the if we use point estimates of $\bm{\mathcal{I}}_k$ in SORF it becomes a special case of EMORF. In particular, the point estimates for ${\mathcal{I}}^i_k\ \forall\ i$ in \cite{chughtai2022outlier} can be obtained with the following criterion

\begin{align}
	{\hat{\mathcal{I}}}^i_k &= 
	\begin{cases}
		1 & \text{if } \bar{\tau}^i_k \leq 0,\\
		\epsilon & \text{if } \bar{\tau}^i_k >0
	\end{cases}\label{Isor}
\end{align}

To deduce ${\hat{\mathcal{I}}}^i_k=1$, the following should hold 
\begin{align}
	\Omega^i_k&\geq1-\Omega^i_k\label{sorf2}\\
	\Omega^i_k&\geq0.5 \label{sorf3}
\end{align}
where $\Omega^i_k$ denotes the posterior probability of ${{\mathcal{I}}}^i_k=1$. Using the expression of $\Omega^i_k$ from \cite{chughtai2022outlier} we can write \eqref{sorf3} as
\begin{align} 
	&\frac {1}{1+{\sqrt {\epsilon }}\left({\frac {1}{\theta ^{i}_{k}}-1}\right){\exp\left({\frac {W^{i,i}_{k}}{2R^{i,i}_{k}} (1-\epsilon)}\right)}}\geq0.5\label{sorf4}
\end{align}
which further leads to 
\begin{align}
	&\bar{\tau}^i_k  \leq 0
\end{align}
where
\begin{align}
	\bar{\tau}^i_k  =&{{\frac {W^{i,i}_{k}}{R^{i,i}_{k}} (1-\epsilon)}}+\ln({\epsilon }) + 2\ln \Big({\frac {1}{\theta ^{i}_{k}}-1}\Big) \label{sorf6}
\end{align}
which can be recognized as the particular case of \eqref{eqn_fl_up10_i} given $\mathbf{R}^{i,-i}_k$ and $\mathbf{R}^{-i,i}_k$ vanish as $\mathbf{R}_k$ is considered to be diagonal. The first term in $\eqref{eqn_fl_up10_i}$ reduces to ${{\frac {W^{i,i}_{k}}{R^{i,i}_{k}} (1-\epsilon)}}$ given $	\bm{\Xi}^{i,i}=\frac{1-\epsilon}{R^{i,i}_k}$ and $\bm{\Xi}^{i,-i}, \bm{\Xi}^{-i,i}$ and $\bm{\Xi}^{-i,-i}$ all reduce to zero. Moreover, the second term in \eqref{eqn_fl_up10_i} also disappears resulting in \eqref{sorf6}. 

\section{Choice of the parameters $\theta^i_{k}$ and $\epsilon$}\label{FifthAppendix}
For EMORF we propose setting the parameters $\theta^i_{k}$ and $\epsilon$ the same as in SORF. Specifically, we suggest choosing a neutral value of 0.5 or an uninformative prior for $\theta^i_{k}$ unless some particular information about outlier statistics is available. The Bayes-Laplace and the maximum entropy approaches for obtaining uninformative prior for a finite-valued parameter lead to the choice of the uniform prior distribution \cite{9931968,turkman2019computational}. Moreover, the selection has been justified in the design of outlier-resistant filters assuming no prior information about the outliers statistics is available \cite{chughtai2022outlier,8869835}. For $\epsilon$ we recommend its value to be close to zero since the exact value of $0$ denies the VB/EM updates as in \cite{chughtai2022outlier}.  
}

\bibliography{main.bib}
\bibliographystyle{IEEEtran}

\end{document}